\documentclass[structabstract]{aa}  
\usepackage{graphicx}
\usepackage{txfonts}
\usepackage{color}
\usepackage[normalem]{ulem}
\usepackage{enumerate}
\usepackage[]{subfigure}
\usepackage{lscape}
\usepackage{rotating}
\usepackage{longtable}
\usepackage{lipsum}
\usepackage{ulem}
\begin{document}
   \title{The variability behavior of CoRoT M-giant Stars
\thanks{
The CoRoT space mission was developed and  is operated by the French
space agency CNES, with the participation of ESA's RSSD and Science Programmes,
Austria, Belgium, Brazil, Germany, and Spain.}}

\author{C. E. Ferreira Lopes \inst{1,2}
\and V. Neves \inst{1}
\and I. C. Le\~ao \inst{1}
\and D. B. de Freitas \inst{1}
\and B. L. Canto Martins \inst{1} 
\and A. D. da Costa \inst{1}
\and F. Paz-Chinch\' on \inst{1}
\and M. L. Das Chagas \inst{1}
\and A. Baglin \inst{3}
\and E. Janot-Pacheco \inst{4}
\and J. R. De Medeiros \inst{1}
}

\authorrunning{C. E. Ferreira Lopes et al.}
\titlerunning{The Variability of M-giant Stars}
   \institute{Departamento de F\'isica, Universidade Federal do Rio Grande do Norte, Natal, RN, 59072-970 Brazil\\
              \email{carlos\_eduardo@dfte.ufrn.br}
   \and SUPA (Scottish Universities Physics Alliance) Wide-Field Astronomy Unit, Institute for Astronomy, School of Physics and Astronomy, University of Edinburgh, Royal Observatory, Blackford Hill, Edinburgh EH9 3HJ, UK 
   \and LESIA, UMR 8109 CNRS, Observatoire de Paris, UVSQ, Universit\'e Paris-Diderot, 5 place J. Janssen, 92195 Meudon, France
   \and Universidade de S\~ao Paulo/IAG-USP, rua do Mat\~ao, 1226, Cidade Universit\'aria, S\~ao Paulo, SP, 05508-900 Brazil           
   }

   \date{Received Month Day, Year; accepted Month Day, Year}


 \abstract
   {For 6 years the Convection, Rotation, and Planetary Transits (CoRoT) space mission has acquired photometric data from more than one hundred thousand point sources towards and directly opposite from the inner and outer regions of the Galaxy. The high temporal resolution of the CoRoT data combined with the wide timespan of the observations has enabled the study of short and long time variations in unprecedented detail.}
   {The aim of this work is the study of the variability and evolutionary behavior of M-giant stars using CoRot data.}
   {From the initial sample of 2534 stars classified as M-giants in the CoRoT databasis, we selected $1428$ targets that exhibit well defined variability, using visual inspection. Then, we defined three catalogs: C1 -- stars with $T_{\mathrm{eff}} < 4200$ K and LCs displaying semi-sinusoidal signatures; C2 -- rotating variable candidates with $T_{\mathrm{eff}} > 4200$ K; C3 -- long period variable candidates (with LCs showing variability period up to the total time span of the observations). The variability period and amplitude of C1 stars were computed using Lomb-Scargle and harmonic fit methods. Finally, we used C1 and C3 stars to study the variability behavior of M-giant stars.}
   {The trends found in the $V-I$ vs $J-K$ color-color diagram are in agreement with standard empirical calibrations for M-giants. The sources located towards the inner regions of the Galaxy are distributed throughout the diagram while the majority of the stars towards the outer regions of the Galaxy are spread between the calibrations of M-giants and the predicted position for Carbon stars. The stars classified as supergiants follow a different sequence from the one found for giant stars. We also performed a KS test of the period and amplitude of stars towards the inner and outer regions of the Galaxy. We obtained a low probability that the two samples come from the same parent distribution. The observed behavior of the period-amplitude and period-effective temperature($T_{\mathrm{eff}}$) diagrams are, in general, in agreement with those found for \emph{Kepler} sources and ground based photometry, with pulsation being the dominant cause responsible for the observed modulation. We also conclude that short-time variations on M-Giant stars do not exist or are very rare and the few cases we found are possibly related to biases or background stars.}
   {}


   \maketitle

\section{Introduction}

The Convection, Rotation, and Planetary Transits \citep[CoRoT;][]{Baglin-2007} and \emph{Kepler} \citep[][]{Borucki-2010} photometry space missions, have enabled us to improve our knowledge of a large variety of stellar phenomena, such as the rotational behavior of different families of stars \citep[e.g.,][]{Affer-2012,De-Medeiros-2013,Nielsen-2013,McQuillan-2013,De-Freitas-2013}, detection and characterization  of a wide variety of extrasolar planets \citep[e.g.,][]{Leger-2009,Borucki-2012,Batalha-2013} and their host stars \citep[e.g.,][]{Brown-2011,Dressing-2013,Morton-2014,Huber-2014}, the analysis of stellar interiors using asteroseismology \citep[e.g.,][]{Huber-2012, Silva-Aguirre-2012,Hekker-2013}; characterization of surface differential rotation \citep[][]{Lanza-2014}; and the detection of eclipsing binary systems \citep[e.g.,][]{Maceroni-2009,Slawson-2011,Maciel-2011}. The aforementioned science cases are only a few major topics among the many aspects of time-domain, high-precision photometry. The data provided by the CoRoT and \emph{Kepler} space missions represent the most complete dataset for the study of stellar variability available to date \citep[e.g.,][]{De-Medeiros-2013,Walkowicz-2013}. In general the results from each mission agree with each other, but there is a lack of a comparative study regarding M-type stars.

One of the fundamental questions in debate in the literature is the evolution and nature of the pulsations in M giants. Our recent understanding of how M giants oscillate was obtained from ground-based surveys such as MACHO, OGLE, and EROS \citep[e.g.][]{Wood-1996,Alard-2001,Lebzelter-2002,Kiss-2003,Soszynski-2007,Riebel-2010,Wisniewski-2011,Soszynski-2013}, as well as solar neighbourhood observations \citep[][]{Tabur-2010}.

Another important question concerns the evolution and nature of the pulsations in the GK-M type giant transition in the RGB.  Whether the main physical mechanism of the pulsations of M giants is self-excited (Mira-type) pulsations or stochastic (Solar-type) oscillations is still not clear \citep[e.g.][]{Dziembowski-2001,Christensen-Dalsgaard-2001,Bedding-2005}. 

In this context, \citet[][]{Tabur-2010} showed that, as we advance through the RGB, the observed low amplitude of stochastically excited solar-type oscillations, typical of the GK-type giants, progress to a mixture of Mira-like and solar-like variability, as found in SR variables, and ending with stable, mono-periodic, Mira-like pulsations.  Afterwards, \citet[][]{Mosser-2013}, studied global oscillation parameters of \emph{Kepler} red giants, concluding that the main excitation mechanism in M Giant SR variables are solar-like oscillations, confirming the findings by \citet[][]{Dziembowski-2010}, but could not disentangle RGB from AGB stars.  \citet[][]{Banyai-2013} performed a detailed study of the variability of M-type giants using \emph{Kepler} data as well. They affirm they could distinguish between solar-like oscillations and larger amplitude pulsations. They found a correlation of solar-like oscillations with period that closely follows the well-known $v_{max}$ amplitude scaling relations \citep[e.g.][]{Mosser-2010,Huber-2011,Mosser-2013}, but  they found a sharp ending to this correlation at $\log P \sim 1$. This feature may be an indication that a different excitation mechanism dominates from this point forward. Does this point mark the transition between solar-like and Mira-type pulsations? In their diagram, the stars classified as having a few periodic components (Miras and SRs) then follow a trend with a steeper slope, for $\log P$ greater than 1. This trend is similar to the one found by \citet[][]{Tabur-2010} for bright M-giant stars, depicting overtone pulsators, with a few stars following the fundamental mode trend (typical of Mira-type stars) as well.

Another point of discussion concerns the observation of  rapid variations of brightness on time-scales between 3 minutes to 30 days in long period variables (hereafter LPVs) \citep[e.g.,][]{Schaefer-1991,Maffei-1995,DeLaverny-1998}. However, these events were not confirmed in recent studies \citep[e.g.,][]{Mais-2004,Wozniak-2004,Lebzelter-2011,Hartig-2014}. Indeed, \citet[][]{Lebzelter-2011} and \citet[][]{Hartig-2014}, using data from CoRoT  and \emph{Kepler}, did not confirm such variations, despite photometric precisions of the order of a few mmag. Therefore, the observed short-time variations are very rare or not physical.

In this study we perform a comprehensive analysis of the variability of M-giants observed by CoRoT, and we compare our results with previous studies regarding the variability behavior of M-giant stars observed by others authors. The paper is structured as follows.  In Sect.~\ref{methods}, we describe our working sample and our catalog. In Sect.~\ref{results}, we present our results and compare them with previous works from the literature. We also investigate the difference between CoRoT stars located towards the inner and outer regions of the Galaxy. Finally, in Sect.~\ref{conclusions}, we present our conclusions and discuss future perspectives.

%
\section{Working Sample}\label{methods}

The CoRoT satellite was launched in 2006. It collected point-source photometric data towards inner (around RA = 18h50m, Dec = +00º) and outer (around RA = 06h50m, Dec = +00º) regions of the Galaxy  over a period of 6 years. A total of $162\!,789$ sources were observed, with 5762 classified as M-type stars \citep[][]{Deleuil-2009}. These stars constitute approximately $4\%$ of all stars observed by CoRoT and are divided in two main groups: giants (luminosity classes I, II, and III)  and dwarfs (luminosity class V). giants comprise approximately $44\%$ of all M-type stars detected by CoRoT, whereas the percentage of M dwarfs is approximately $56\%$. A total of 973 M-type stars were observed at least twice, providing a unique sample for studying long-term photometric variations. Table \ref{alldata} lists the number of M-type stars observed with CoRoT towards the inner (center) and outer (anticenter) regions of the Galaxy.

\begin{table}[h]
  \centering
  \begin{tabular}{c|c|c|c|c}
  \hline
  \hline
   CoRoT Region & I & II & III & V \\
\hline
Center     &  136 & 189 & 1339 & 1807  \\ 
Anticenter &   48 &  31 &  791 & 1421 \\ 
\hline
\end{tabular}
\caption{ CoRoT M-type stars taken from EXO-Dat \citep[][]{Deleuil-2009} distributed by luminosity class.}
\label{alldata}	
\end{table}

The latest CoRoT data release N2\footnote{http://idoc-corot.ias.u-psud.fr/} provides light curves (LCs) that are corrected for instrumental noise cause by electronic, background and jitter sources. The effect of the South Atlantic Anomaly (SAA) passage was also included, and hot pixels of the detector are now flagged \citep[][]{Auvergne-2009}. At the time of the beginning of this work, the data were not yet completely corrected and some problems remained in the released data, requiring further treatment before the data was ready for analysis. In particular, the data contained long-term trends produced by CCD temperature variations, jumps (discontinuities) produced by hot pixels, and outliers \citep[][]{Auvergne-2009}. The post-processing of the LCs performed in different works follows different methods, according to the objectives of the works \citep[e.g.,][]{Renner-2008,Basri-2011,Affer-2012,De-Medeiros-2013}, and thus far, there is no standard method to analysed the processed data. In the present work, we followed the general guidelines presented by \citet[][]{De-Medeiros-2013}.

First, we selected all sources classified as M-type giant stars in the CoRoT database. This first pre-selection, which yielded $2534$ objects, provided us with a reasonable  of actual giants, although the luminosity class specified in the CoRoT database is sometimes incorrect. 

We also obtained the Harris $V$ and Sloan-Gunn $r'$ and $i'$ photometry from the Exo-Dat catalog \citep[][]{Deleuil-2009} and $JHK_{s}$ photometry from the Two Micron All Sky Survey \citep[\textit{2MASS}-][]{Skrutskie-2006} catalog. The indexes $V$, $r'$, and $i'$ were transformed into Johnson $V$ and $I$-band photometry using the relations in \citet{Deleuil-2009}, with propagation of the respective errors.

Then, we derredened the photometry using the extinction coefficients $A_{V}$, $A_{I}$, $A_J$, and $A_{K}$ calculated with the relations from \citet[][]{Schlafly-2011} and $E(B-V)$ reddening maps taken from \citet{Schlafly-2014}. We took the uncertainties from the coefficients and reddening maps and propagated them to obtain the final uncertainty of the corrected photometry. The transformed Johnson photometry and the unreddened photometry are listed in Tables \ref{tab-cat01} and \ref{tab-cat02}.

\subsection{Cross-identification}\label{cross}

We performed a systematic cross-check of the 2534 M-giants to identify previous studies in the literature. To this end, we used the SIMBAD database\footnote{http://simbad.u-strasbg.fr/simbad/}, the General Catalog of Variable Stars (Samus et al. 2012), the AAVSO International Variable Star Index (VSX v1.1, now including 325,346 variable stars; \citet[][]{Watson-2014}), the New Catalog of Suspected Variable Stars \citep[][]{Kazarovets-1998}, the Northern Sky Variability Survey catalog (NSVS; Hoffman et al. 2009), and a few catalogs of carbon stars \citep[][]{MacConnell-1988,Stephenson-1989,Stephenson-1996,Alksnis-2001,Chen-2012,Zacharias-2013}. We also cross-checked with previous studies using CoRoT data \citep[][]{Lebzelter-2011,Sebastian-2012,Guenther-2012,Sarro-2013,De-Medeiros-2013}. Finally, we also searched in many other databases incorporated in the International Virtual Observatory Alliance\footnote{http://www.ivoa.net/}, using  the Tool for OPerations on Catalogues And Tables (TOPCAT)\footnote{http://www.star.bris.ac.uk/~mbt/topcat/}. The search assumed a positional accuracy of 2'' in the sky coordinates.

A total of 465 stars were in common with other studies. Among them there are 390 semi-sinusoidal variables, 45 LPVs, 14 Orion type variable stars, 8 Carbon stars, 6 Mira type stars, 4 binary systems, 2 pre-main sequence stars, 2 high proper-motion stars, 1 Delta Scuti star, and 1 Gamma-Doradus star. Some of these stars have more than one classification. For instance, 15 semi-sinusoidal variables are also classified as LPVs. At the same time, not every star has a known period. The Orion type, Gamma-Doradus, and pre main-sequence stars, as well as stars in binary systems were withdrawn from our sample. The remaining sources were used to compare our period measurements (see Sect.~\ref{submeth}). The stars in common with the literature depicting their variability types are presented in the column 2 of Tables~\ref{tab-cat01} and \ref{tab-cat02}).

\subsection{Effective temperature estimation}\label{physpar}

The $T_{\mathrm{eff}}$ values of the M\textbf{-}giants were calculated using the photometric--effective temperature relation of \citet[][]{van-Belle-1999}, 

\begin{equation}
\label{eq:teff}
T_{\mathrm{eff}} = 3030 + 4750\times 10^{-0.187(V-K)}~\textrm{K}.
\end{equation}

Such a procedure was chosen to provide a homogeneous set of $T_{\mathrm{eff}}$ estimations. This is a precise tool directly based on interferometric measurements. 

The effective temperature was calculated using the unreddened V- and K-band photometry. We transformed the $K_{S}$ magnitudes to $K_{CIT}$ photometry  using the relation of \citet[][]{Carpenter-2001} and calculated the $T_{\mathrm{eff}}$ values. The $K_{CIT}$ photometry was used because it is the closest band to the one that was used to establish the original calibration (van Belle, private communication).

The uncertainties in the $T_{\mathrm{eff}}$ values were estimated by propagating the errors of the corrected magnitudes and the $K_{S}$ transformation uncertainty. Then, we added the photometric uncertainty in quadrature to the estimated errors from Table 8 of \citet{van-Belle-1999}, taking the maximum error of the two $V-K$ bins that were closest to each parameter value. The results are presented in Table~\ref{tab-cat01}.

%
\subsection{Variability analysis}\label{submeth}

A Lomb-Scargle periodogram \citep[][]{Lomb-1976,Scargle-1982} was computed for each LC, except for the LPV candidates (C3 - see Sect. \ref{sec_cat}). We set the low-frequency limit ($f_0$) for each periodogram to be $f_0 = 1/T_{\rm tot}$, where $T_{\rm tot}$ is the total time spanned by the LC. The high-frequency limit was fixed to $f_{N} = 10$~d$^{-1}$, and the periodogram size was scaled to $10^{4}$ elements. The highest periodogram peak, which is referred to as frequency $f_1$ or period $P_1$, was refined following \citet{De-Medeiros-2013}, namely, by maximizing the ratio of the variability amplitude to the minimum dispersion given by \citet[][]{Dworetsky-1983}. Next, the refined frequency $f_1$ was used to calculate a harmonic fit with two harmonics. This fit was used as a model to compute the mean variability amplitude (A) in units of magnitude (mag), and is defined as

\begin{equation}
y(t) = \sum_{j = 1}^{2}\left[ a_{j}\sin\left(2\pi f_{1} j t \right) + b_{j}\cos\left(2\pi f_{1} j t \right) \right] + b_0,
\label{eq_best_harm}
\end{equation}

\noindent in the phase diagram of the main period, where $f_{1}$ is the frequency, $a_{j}$ and $b_{j}$ are the Fourier coefficients, $t$ is the time and $b_0$ is the background level. 

For our M giants (typically stars with $T_{\mathrm{eff}} < 4200$ K - see Sect. \ref{sec_cat}), we defined the variability amplitude as the half of the difference between the maximum and minimum of Eq.~\ref{eq_best_harm} in the phase diagram of the main period, which is similar to the amplitude definitions used by \citet[][]{De-Medeiros-2013} and \citet[][]{Banyai-2013}. This fit was then applied to the LC domain and was subtracted from the time series (prewhitening), thereby yielding a new Lomb-Scargle periodogram. The procedure was iterated $N$ times; in each iteration, the LCs were fitted with a harmonic fit using two harmonics. For each LC, we selected all the independent periods that exhibited a significance level greater than $99\%$, i.e., a false alarm probability (FAP) less than $0.01$. Therefore, a collective set of independent periods for each target was retained to perform an analysis that was similar to that of \citet[][]{Tabur-2010} and \citet{Banyai-2013} (see results presented in Sect.~\ref{results}). The FAP was computed based on the \citet[][]{Eaton-1995} method, which consists of randomizing the temporal bins of the original LC and computing the resulting power spectra \citep[see also][]{Affer-2012}. We produced 1000 modified LCs for each period by randomizing the positions of blocks of adjacent temporal bins, with a block length of $12$ hours, in accordance with \citet{Affer-2012}. Regarding our semi-sinusoidal sources with $T_{\mathrm{eff}} > 4200$ K (C2 - see Sect. \ref{sec_cat}), we performed the analysis once and keep only one period and amplitude (see Table~\ref{tab-cat02}). This process was iterated N times for C1 stars, considering the first three periods above the FAP, to study the multi-modes which are displayed in Table~\ref{tab-cat01}, where we present the photometric information for the LPV candidates (see Table~\ref{tab-cat03}).

\begin{figure*}[!htb] 
\begin{center}

\includegraphics[width=5.75cm,height=4.200cm]{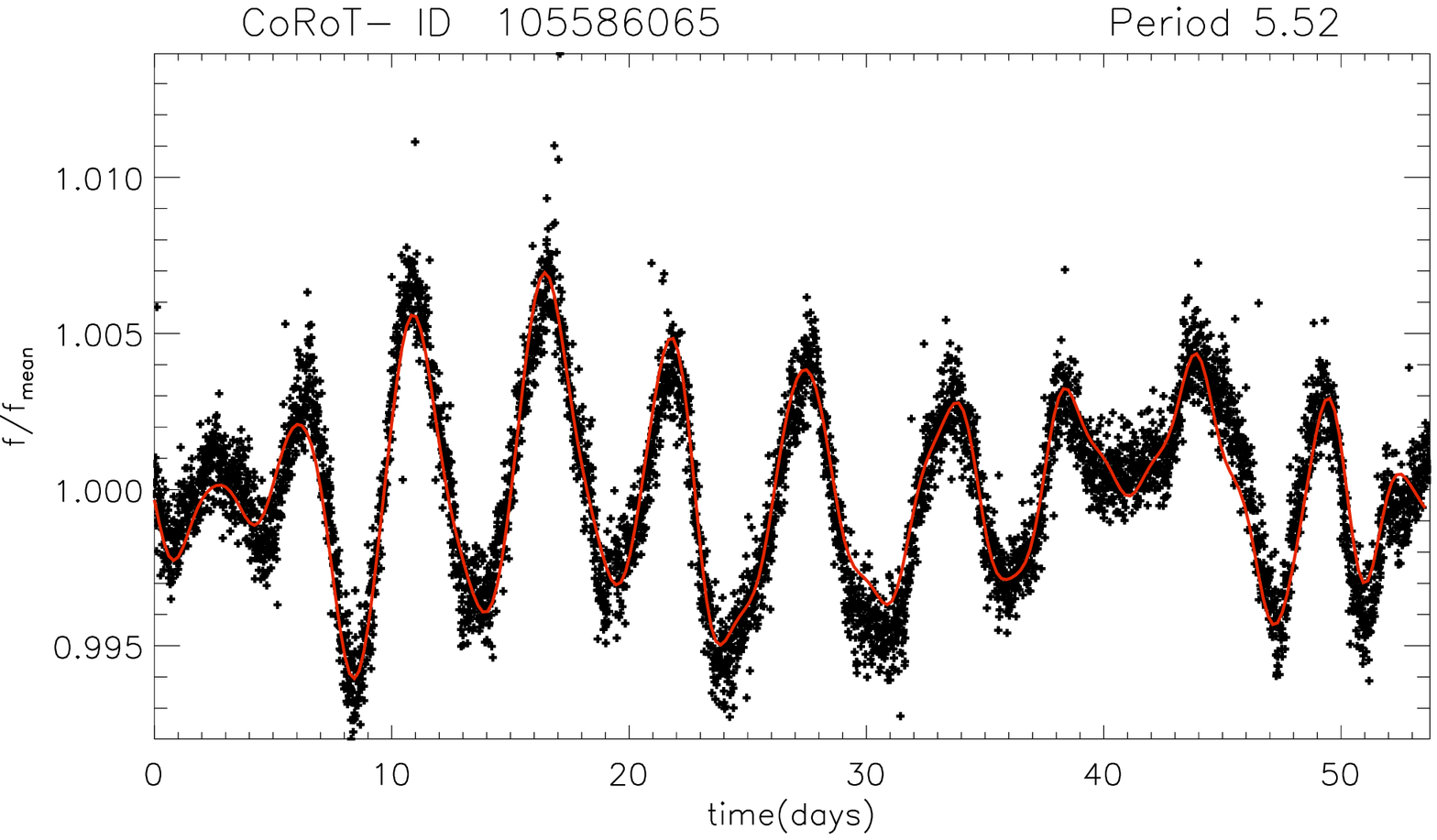}
\includegraphics[width=5.75cm,height=4.200cm]{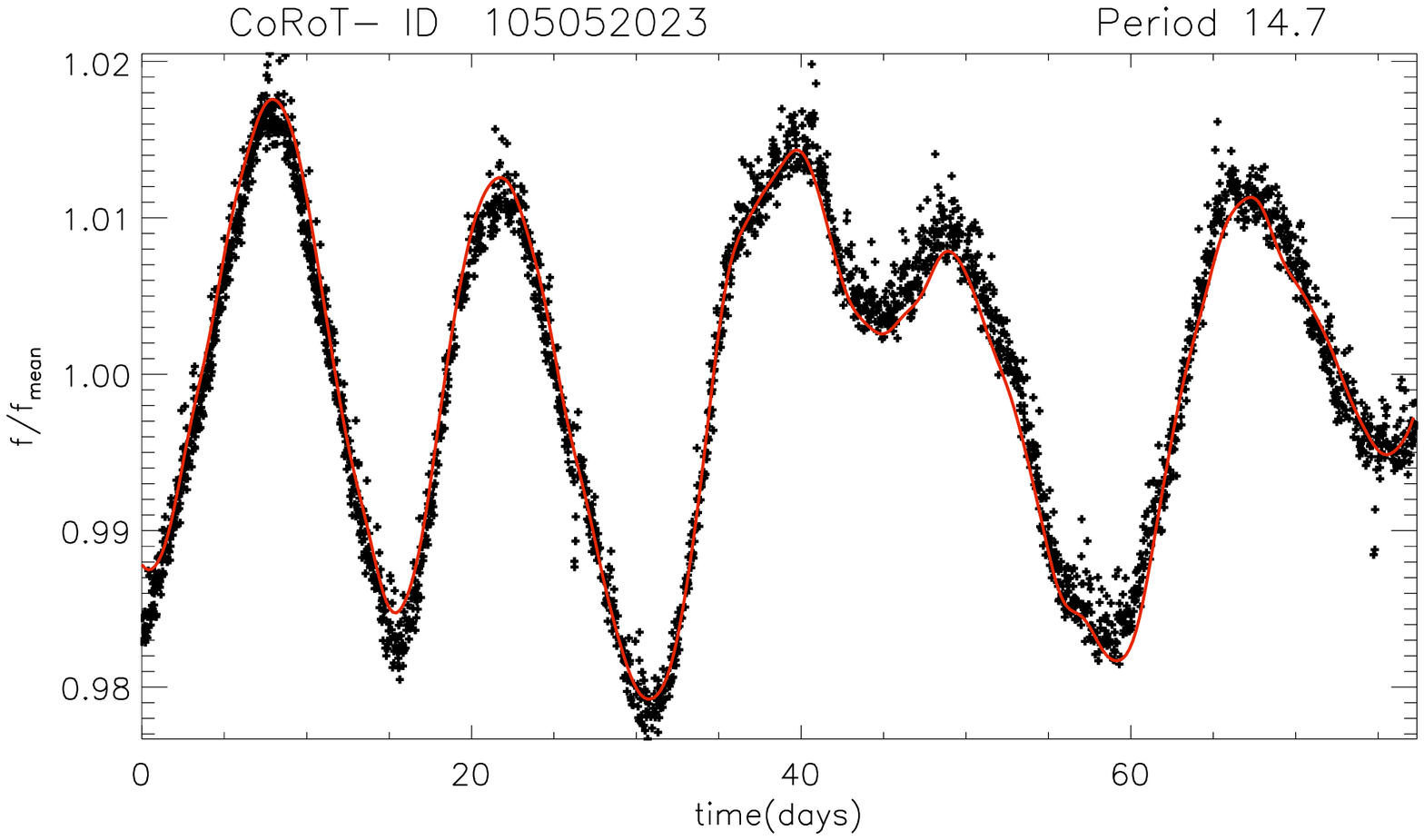}
\includegraphics[width=5.75cm,height=4.200cm]{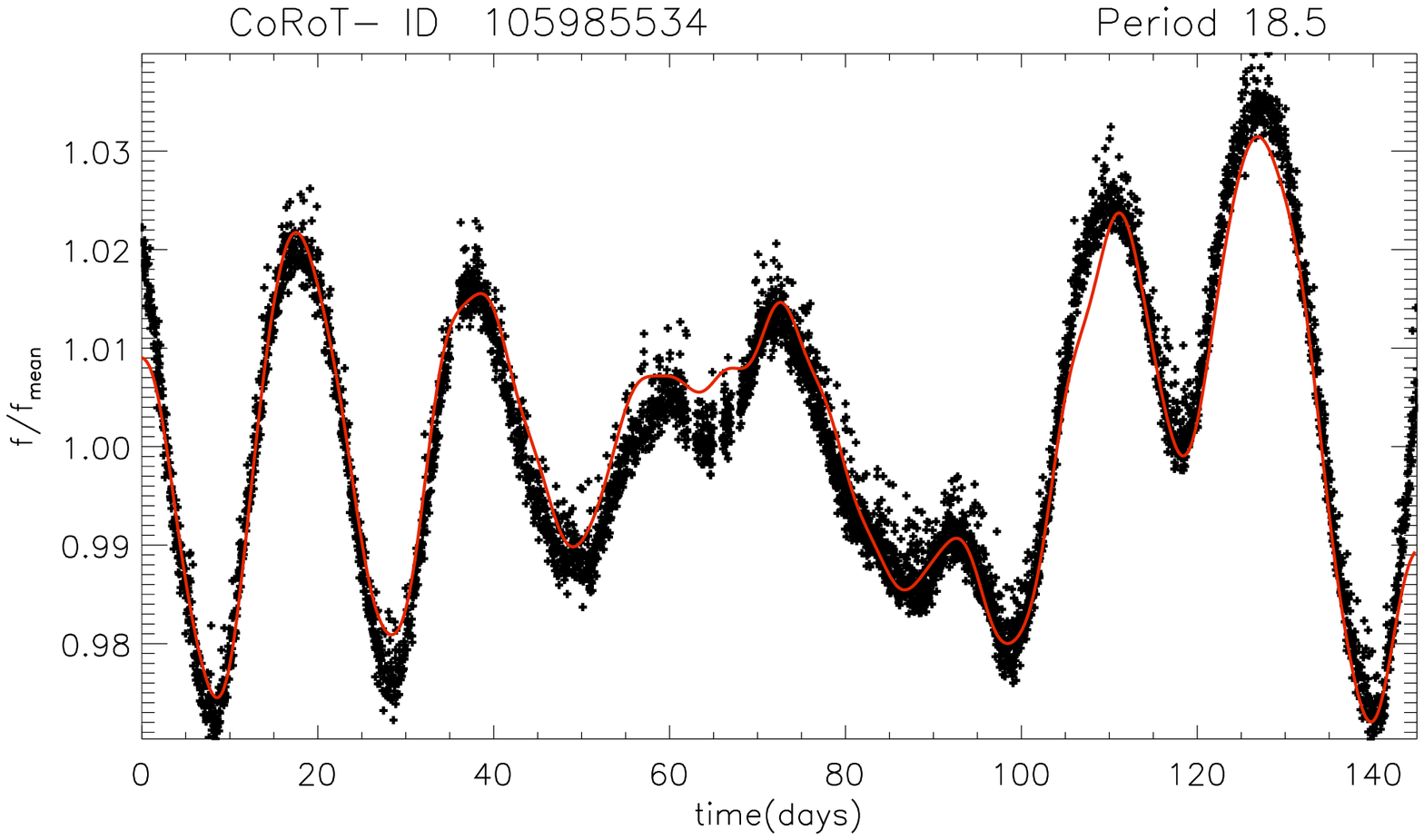}

\includegraphics[width=5.75cm,height=4.200cm]{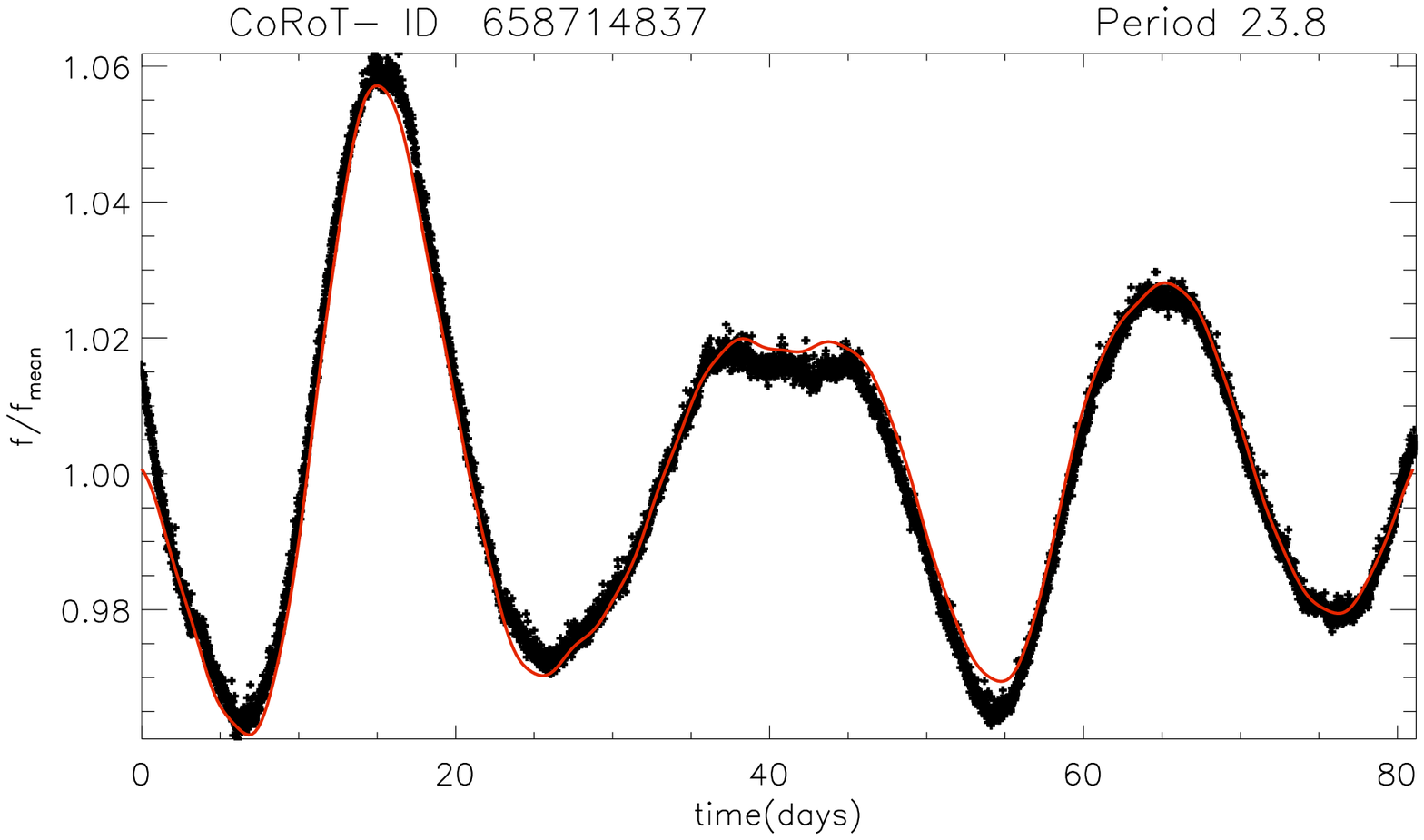}
\includegraphics[width=5.75cm,height=4.200cm]{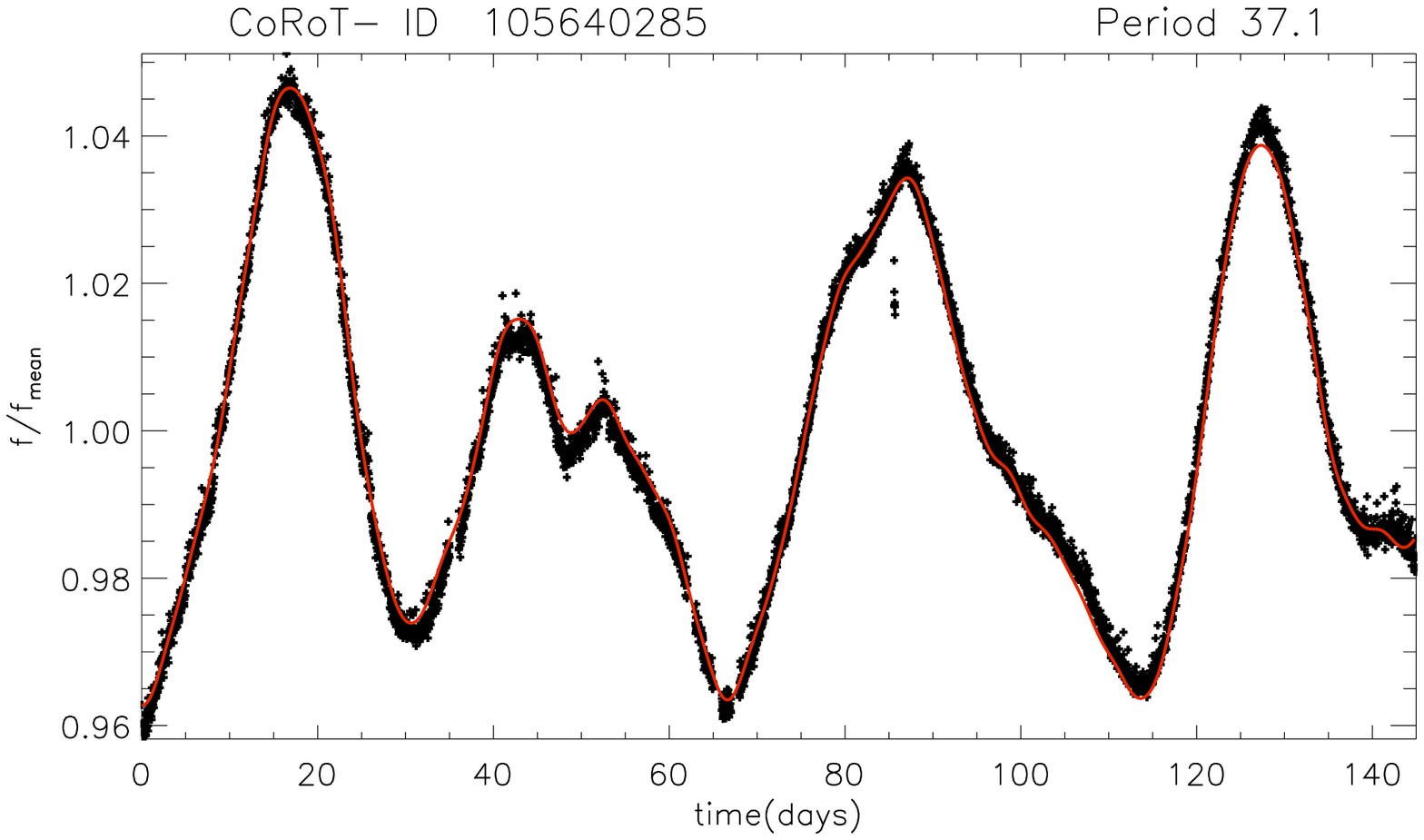}
\includegraphics[width=5.75cm,height=4.200cm]{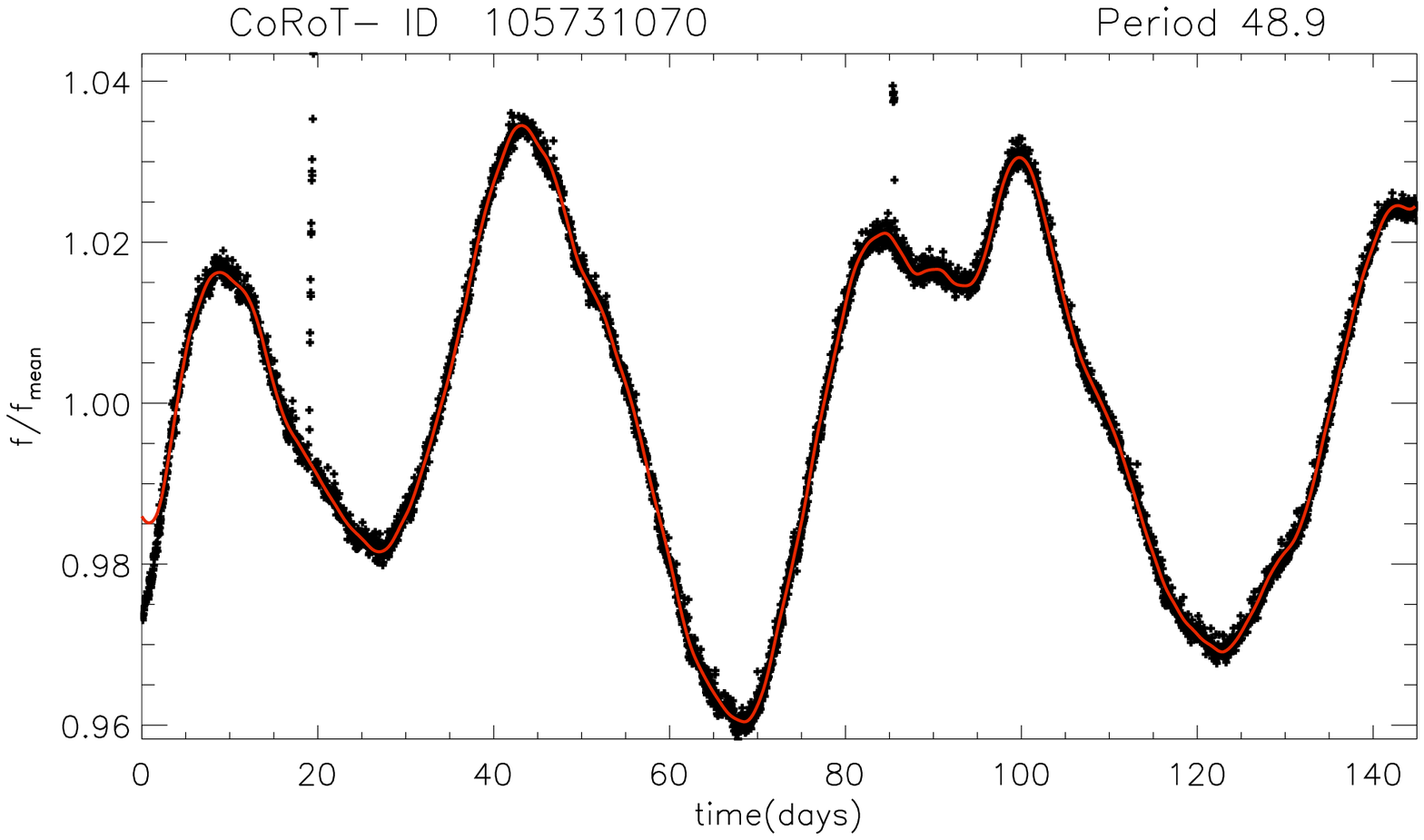}

\includegraphics[width=5.75cm,height=4.200cm]{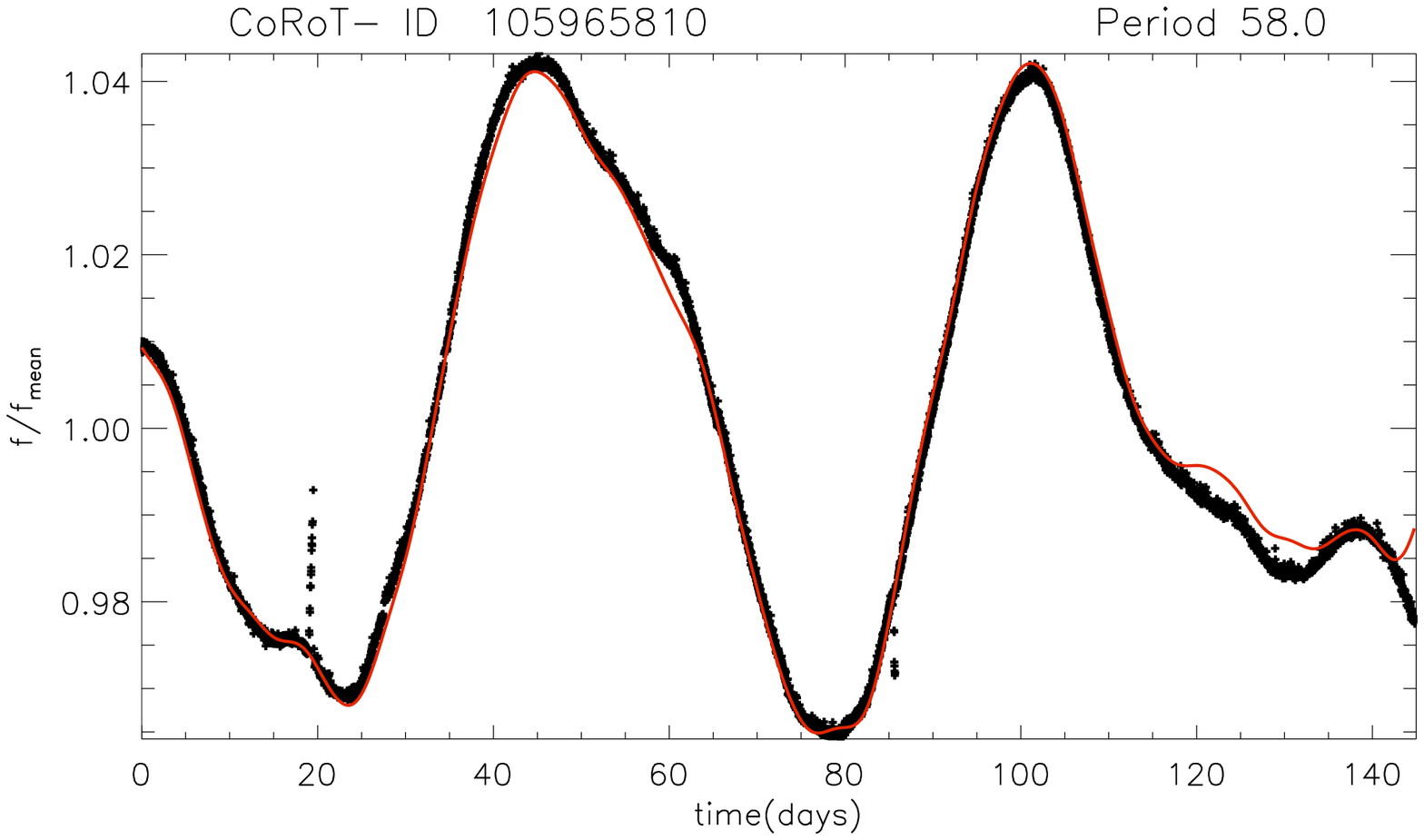}
\includegraphics[width=5.75cm,height=4.200cm]{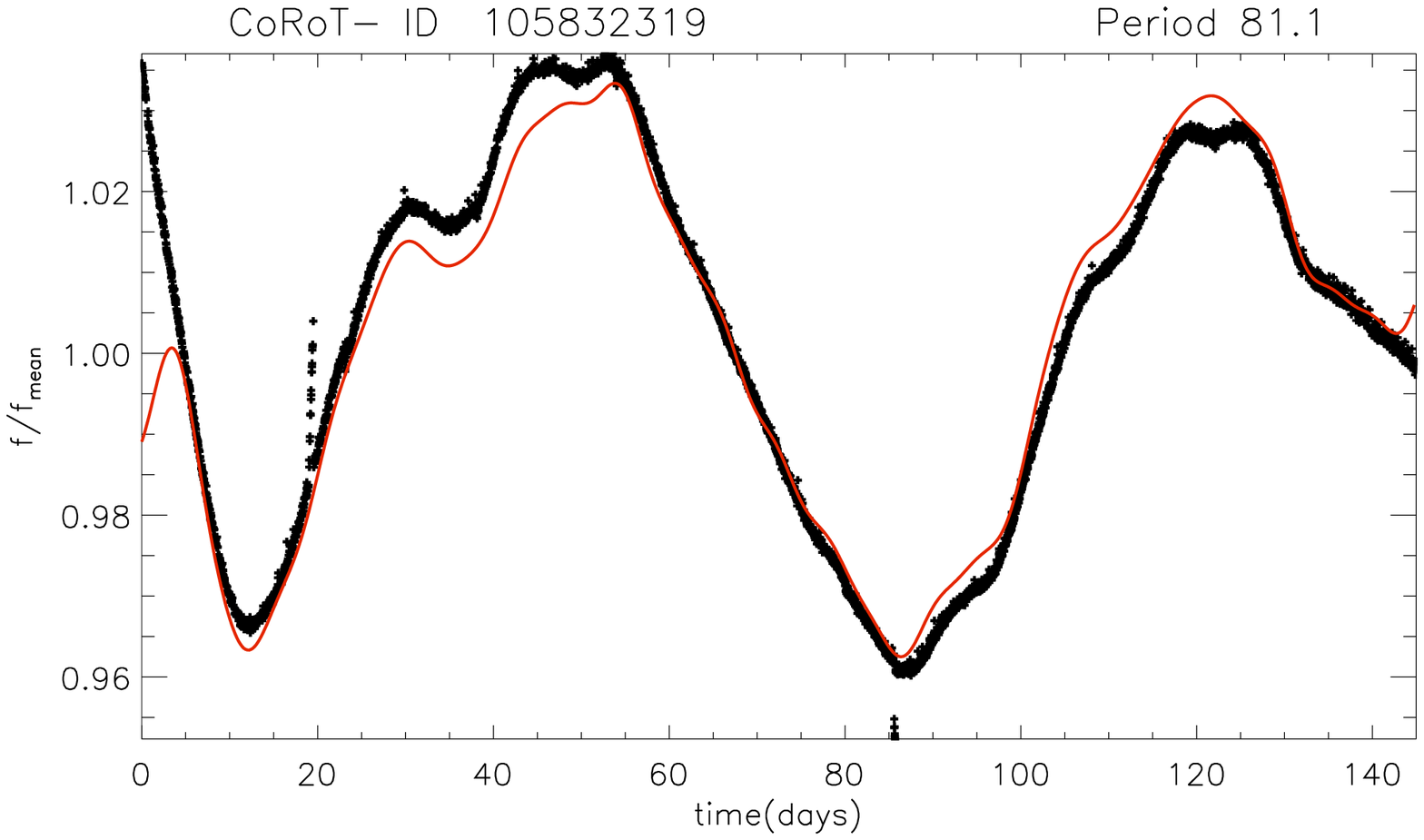}
\includegraphics[width=5.75cm,height=4.200cm]{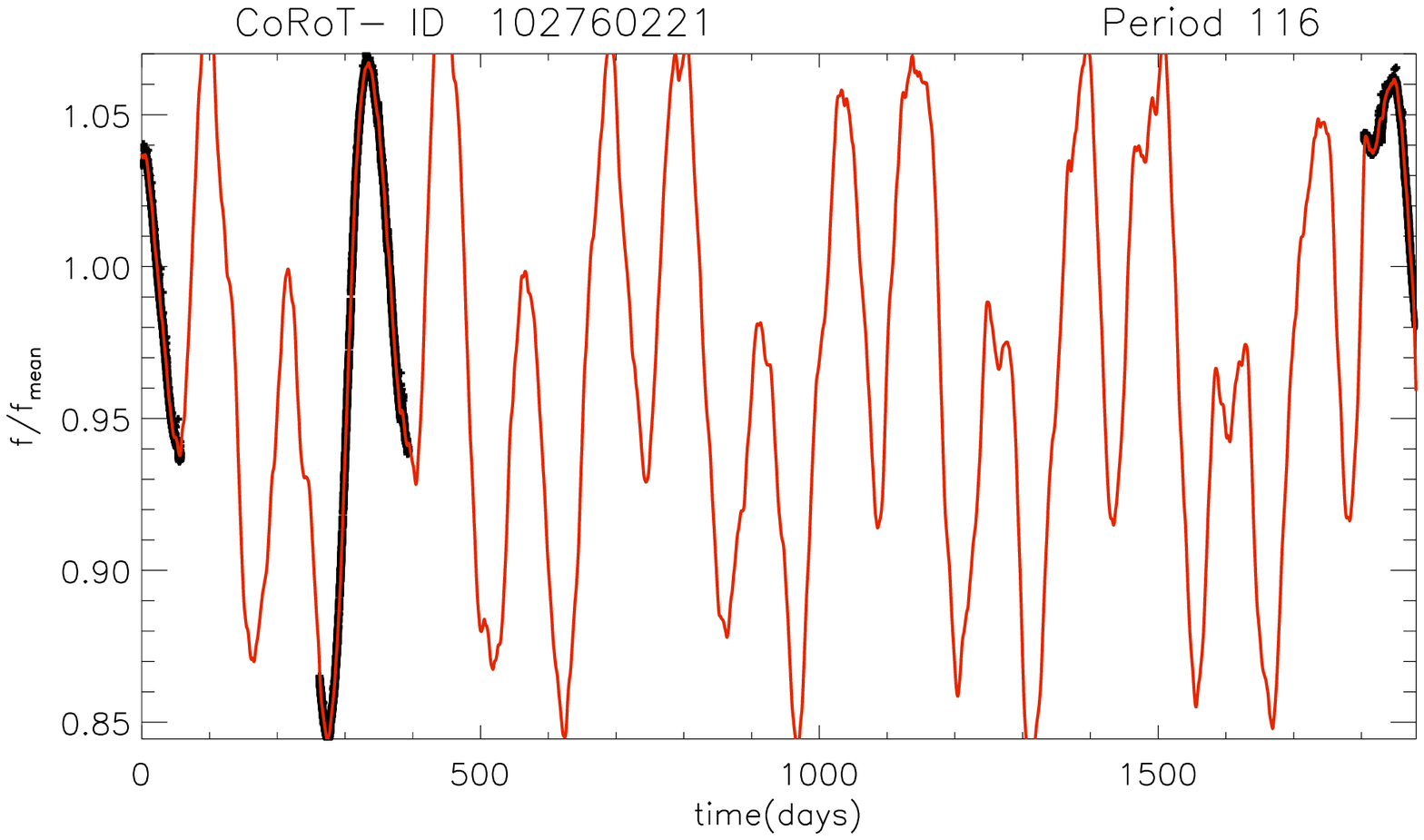}

\includegraphics[width=5.75cm,height=4.200cm]{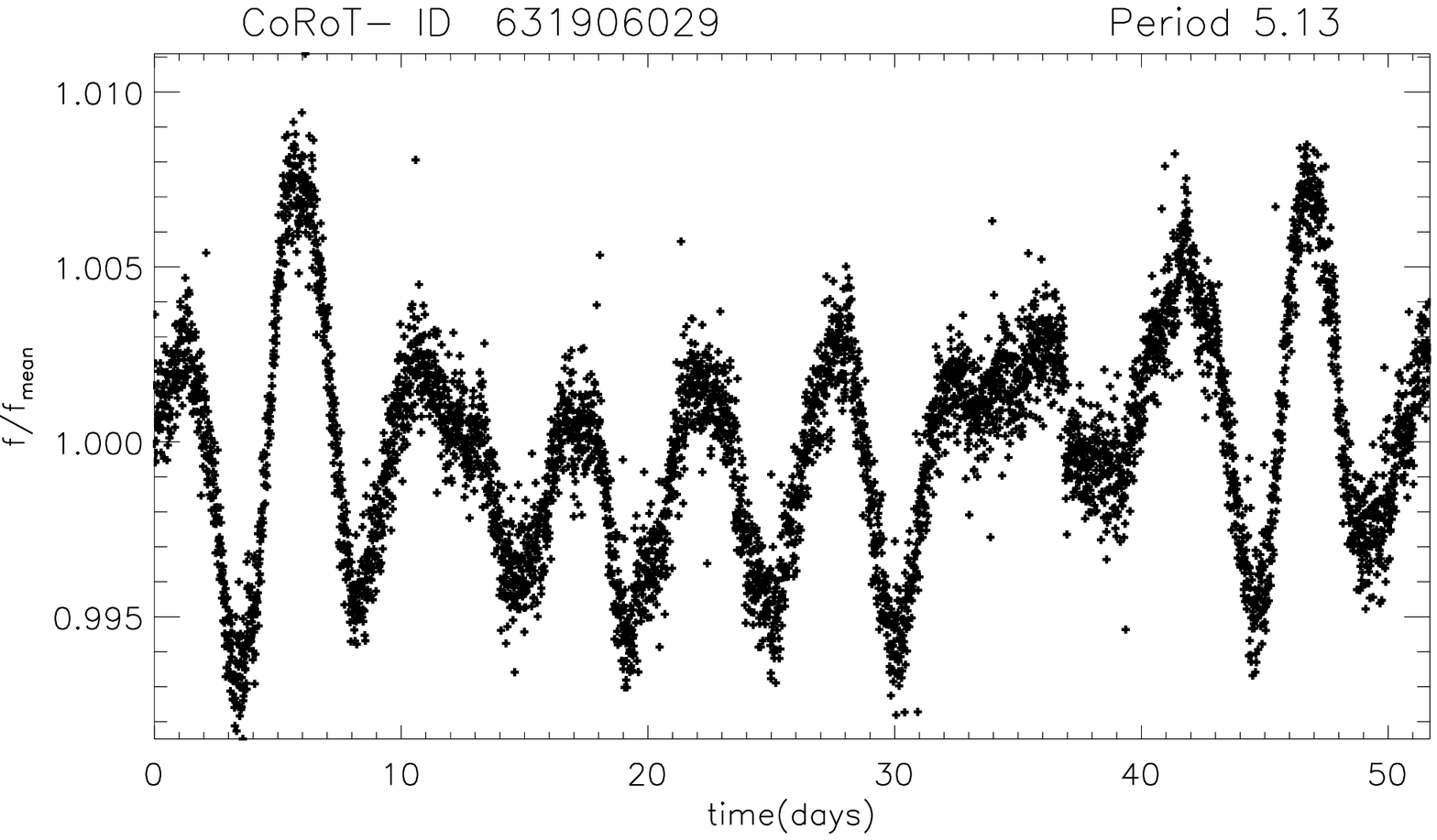}
\includegraphics[width=5.75cm,height=4.200cm]{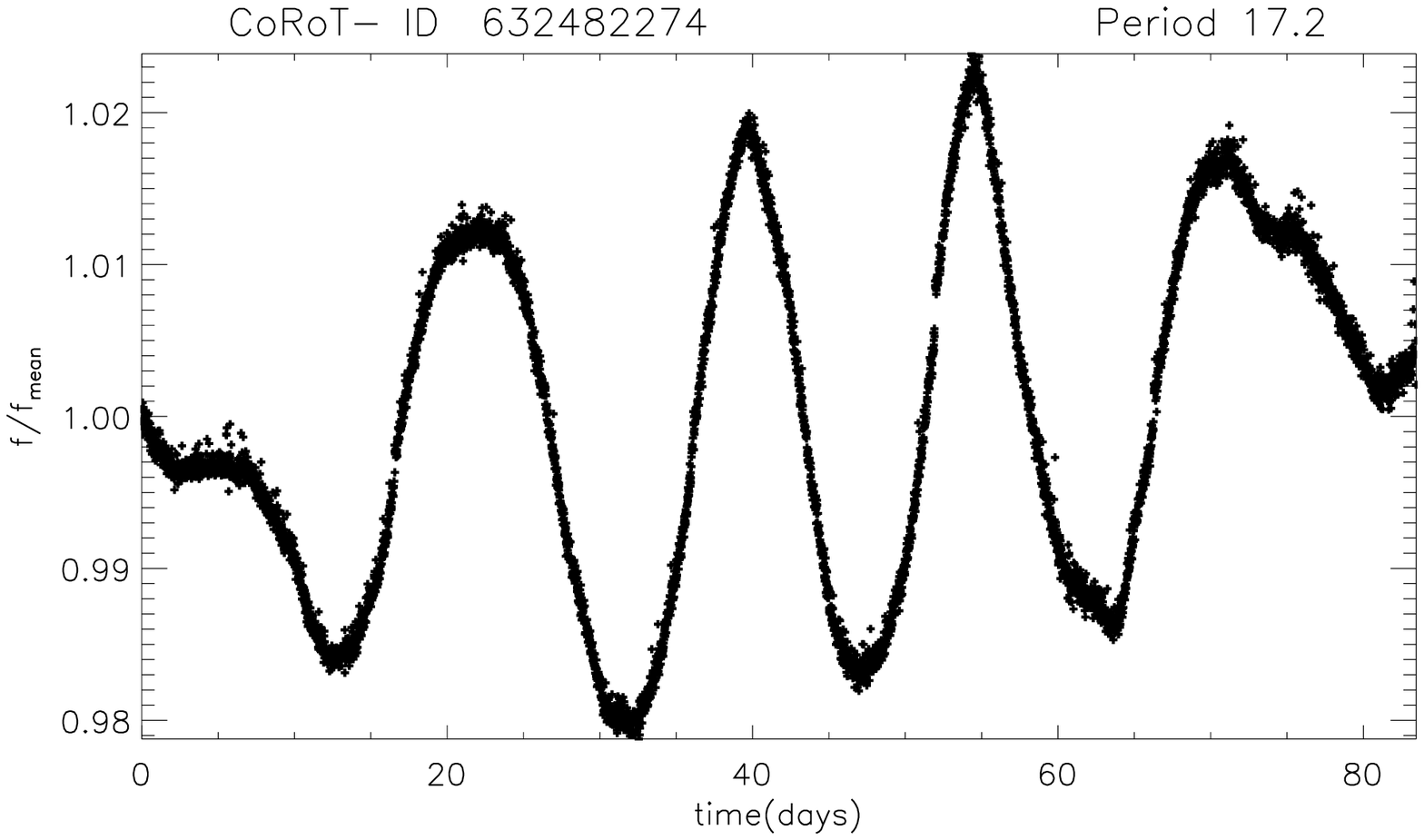}
\includegraphics[width=5.75cm,height=4.200cm]{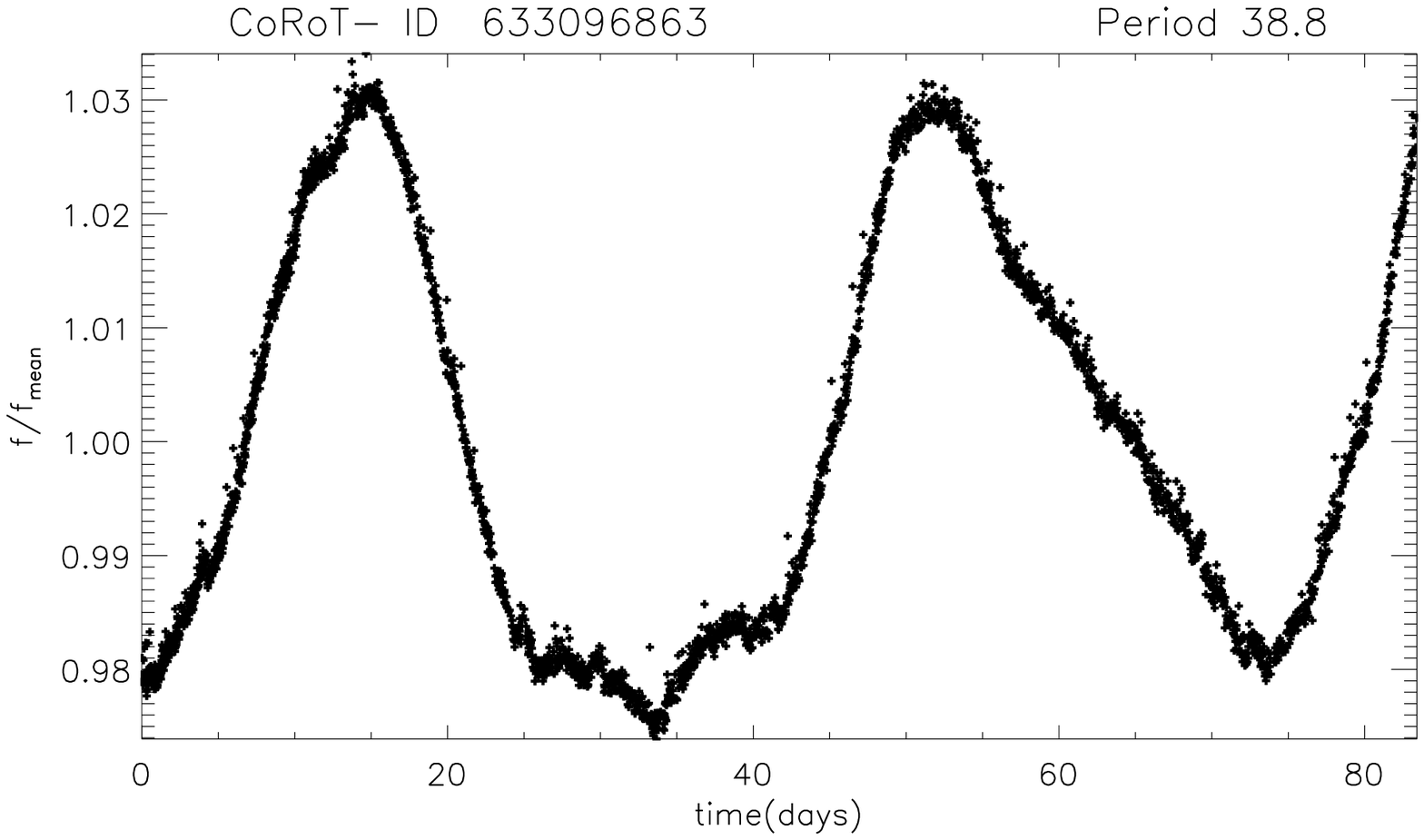}

\includegraphics[width=5.75cm,height=4.200cm]{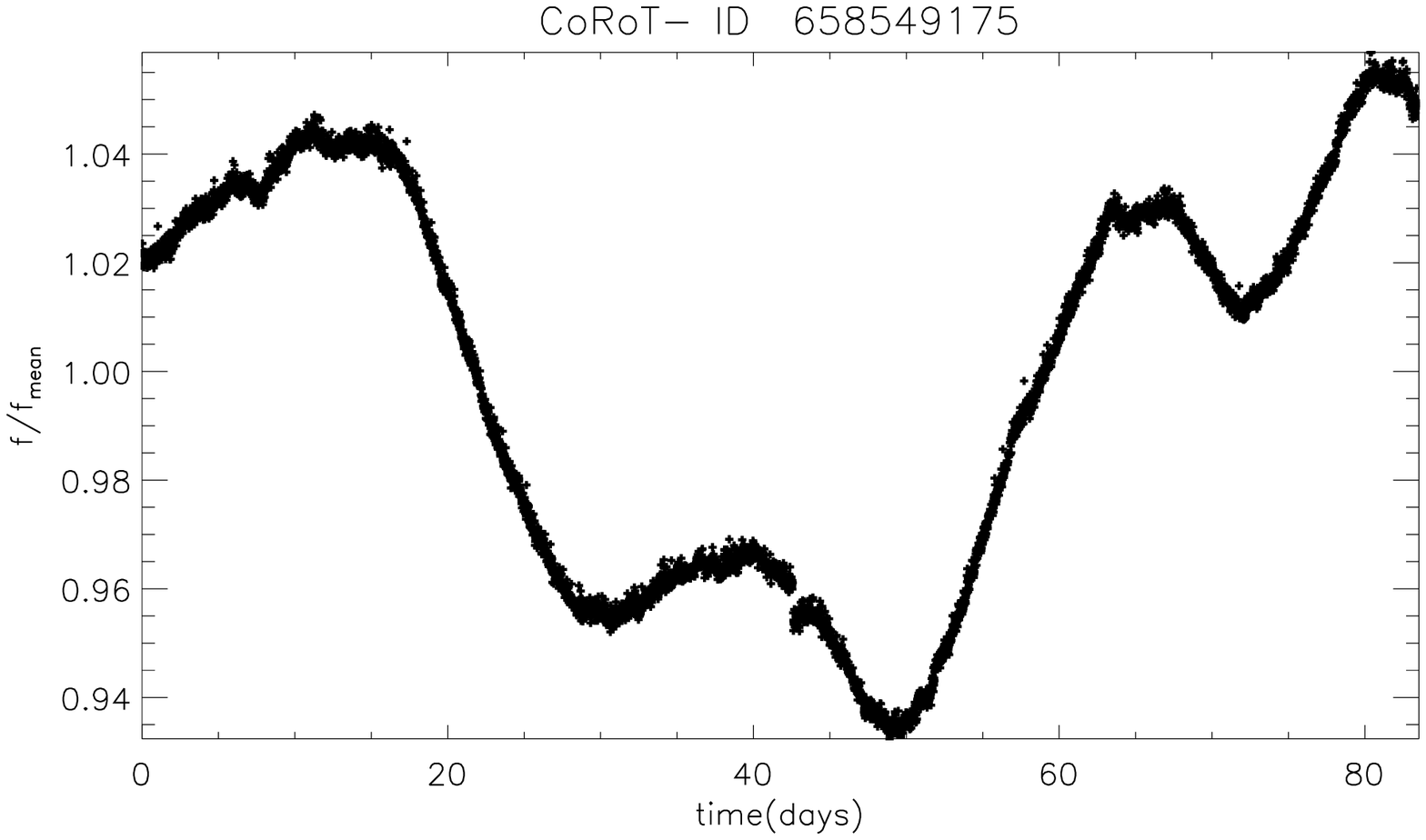}
\includegraphics[width=5.75cm,height=4.200cm]{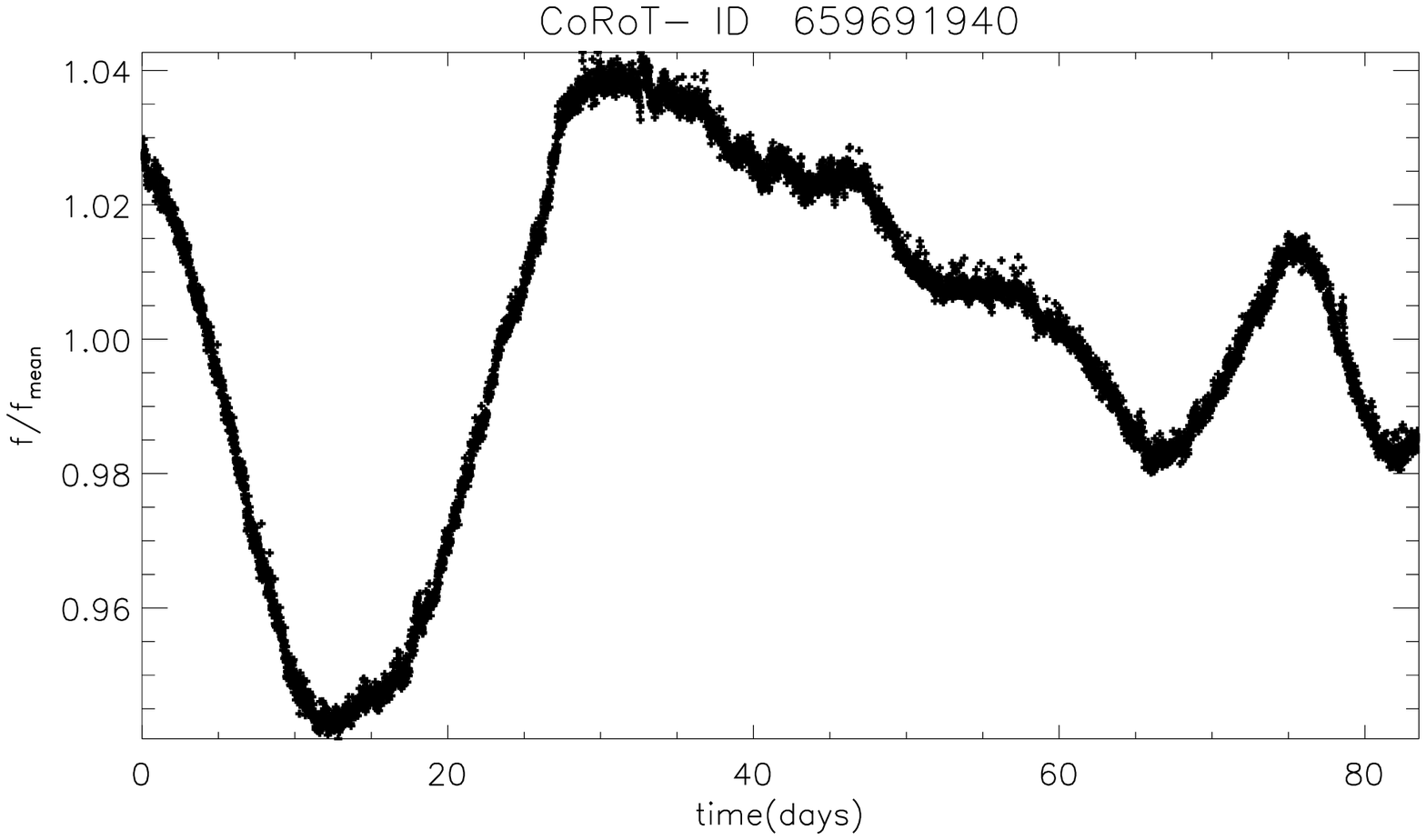}
\includegraphics[width=5.75cm,height=4.200cm]{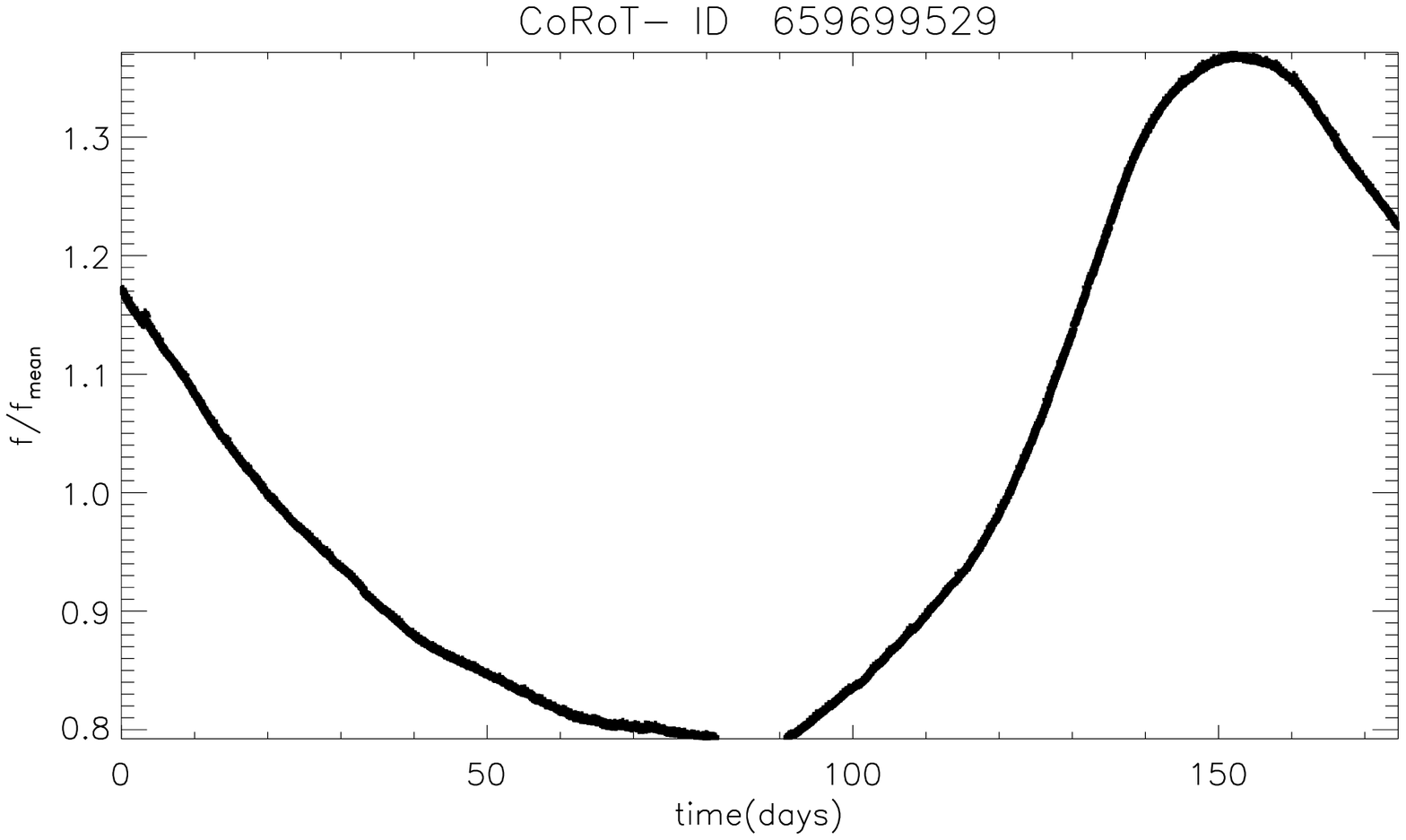}

\caption{Typical LCs of stars from our catalog. CoRoT-ID and period of the C1 (row 1-3) and C2 (row 4) catalogs are shown in the headers. For the C3 (LPVs - row 5) sources only the CoRoT-ID is shown. The red line marks the harmonic fit models of the C1 stars given by Eq.~\ref{eq_best_harm}.}
\label{lc_c1}
\end{center}
\end{figure*}

A total of 434 stars with previously determined periods \citep[][]{Lebzelter-2011,De-Medeiros-2013} were found. For approximately $96\%$ of the targets, the differences between our periods and the other authors are less than $10\%$. We attribute the differences to four main reasons: first, the pipeline improvements made by the CoRoT team now result in better data for analysis; second, the combination of all observations provides a wider time window and thus more precise periods; third, detrending of long-term variations may misshape a small portion of the LCs; and fourth, the first harmonic may be falsely identified as the main frequency when there are too few cycles, as explained in \citet[][]{De-Medeiros-2013}. The precision of the analysis of the nature of the variability, period, and amplitude increases with the number of cycles (the ratio of total time span to the variability period), as discussed in \citet{De-Medeiros-2013}. According to those authors, for the CoRoT LCs, the variability periods with more than three cycles have a confidence level of greater than $80\%$. The photometric instrumental jumps in the CoRoT data  also hinder the analysis of potential long-period variability.

\subsection{The CoRoT M stars variable catalogue}\label{sec_cat}

Our final sample is composed of all semi-sinusoidal variable stars and long period variables candidates in the CoRoT database that were classified as M-giants ($1428$ stars). This catalogue is composed of three main groups: A - $1173$ semi-sinusoidal variables with $T_{\mathrm{eff}} \leq 4200$ K; B - $141$ semi-sinusoidal variables with $T_{\mathrm{eff}} > 4200$ K; C - 114 LPVs candidates (from which 105 have $T_{\mathrm{eff}} \leq 4200$ K, 4 have $T_{\mathrm{eff}} > 4200$ K, and 5 do not have either $V$ or $K$-band photometry). The cutoff of $T_{\mathrm{eff}} < 4200$ K was performed to exclude targets that could be misclassified as M-type.

Fig.~\ref{lc_c1} presents LCs with typical C1 (rows 1-3), C2 (row 4), and C3 (last row) signatures. CoRoT-102760221 (right panel of the third row) shows a combination of CoRoT runs that provides a wider time window, used to search for longer-periods variations. Tables~\ref{tab-cat01}, \ref{tab-cat02}, and \ref{tab-cat03} present the properties of the C1, C2, and C3 sub-catalogs, respectively. Columns 1 to 10 depict the CoRoT ID, right ascension declination, time window of the CoRoT run, spectral type, luminosity class, B, V, J, H, and K$_s$ magnitudes. Columns 11 to 13 show the computed periods, variability amplitudes, and $T_{\mathrm{eff}}$ values, for the stars contained in Tables~\ref{tab-cat01} and \ref{tab-cat02}. The period distribution ranges from $\sim$2 to $\sim$150 day, with a maximum around $\sim 17$~day. The variability amplitudes range from $\sim 1$ to $\sim 900$~mmag, with a maximum of approximately $\sim 10$~mmag.

\begin{figure}[htb]
\begin{center}
\subfigure{\includegraphics[height=7.5cm]{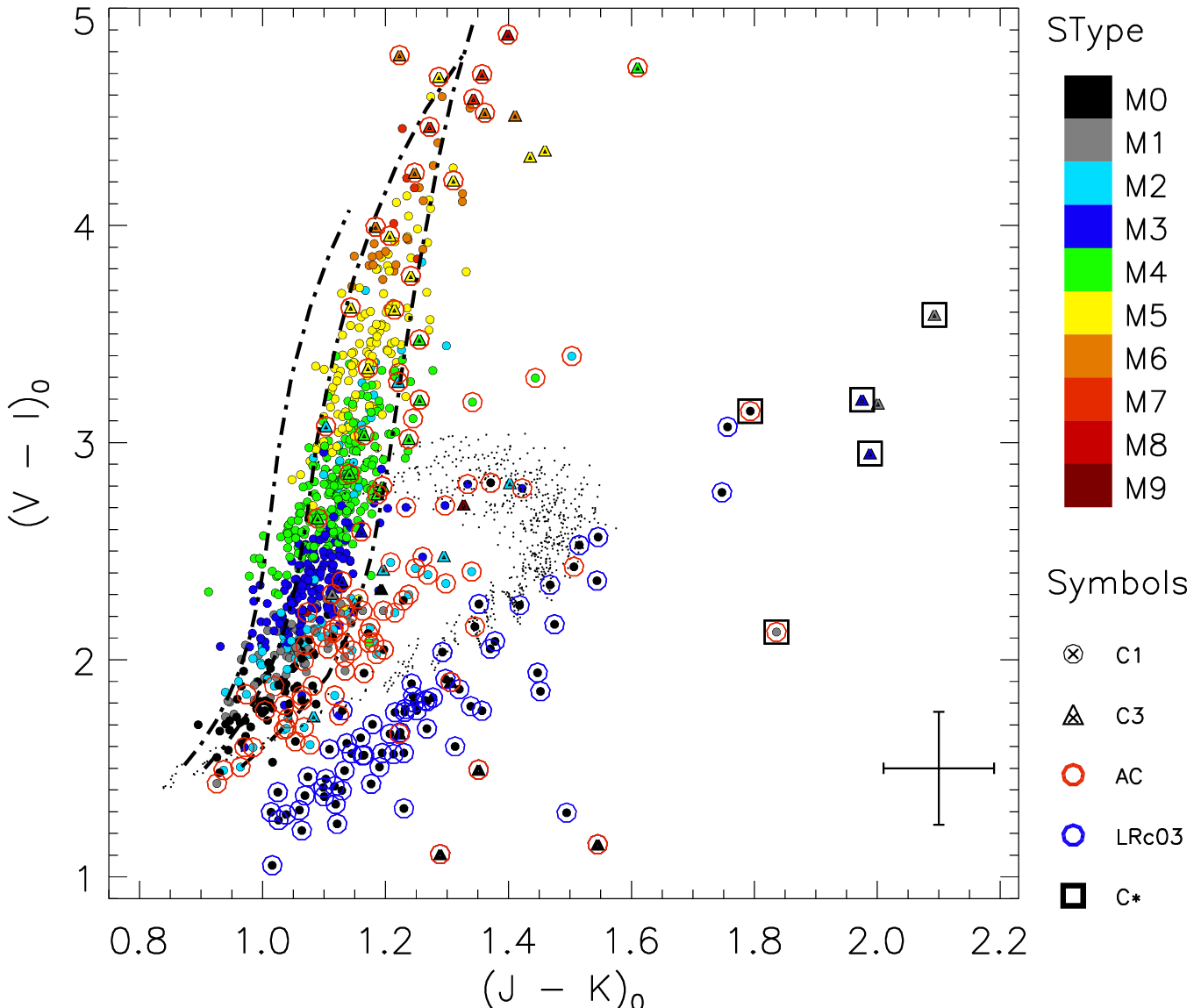}}
\subfigure{\includegraphics[height=7.5cm]{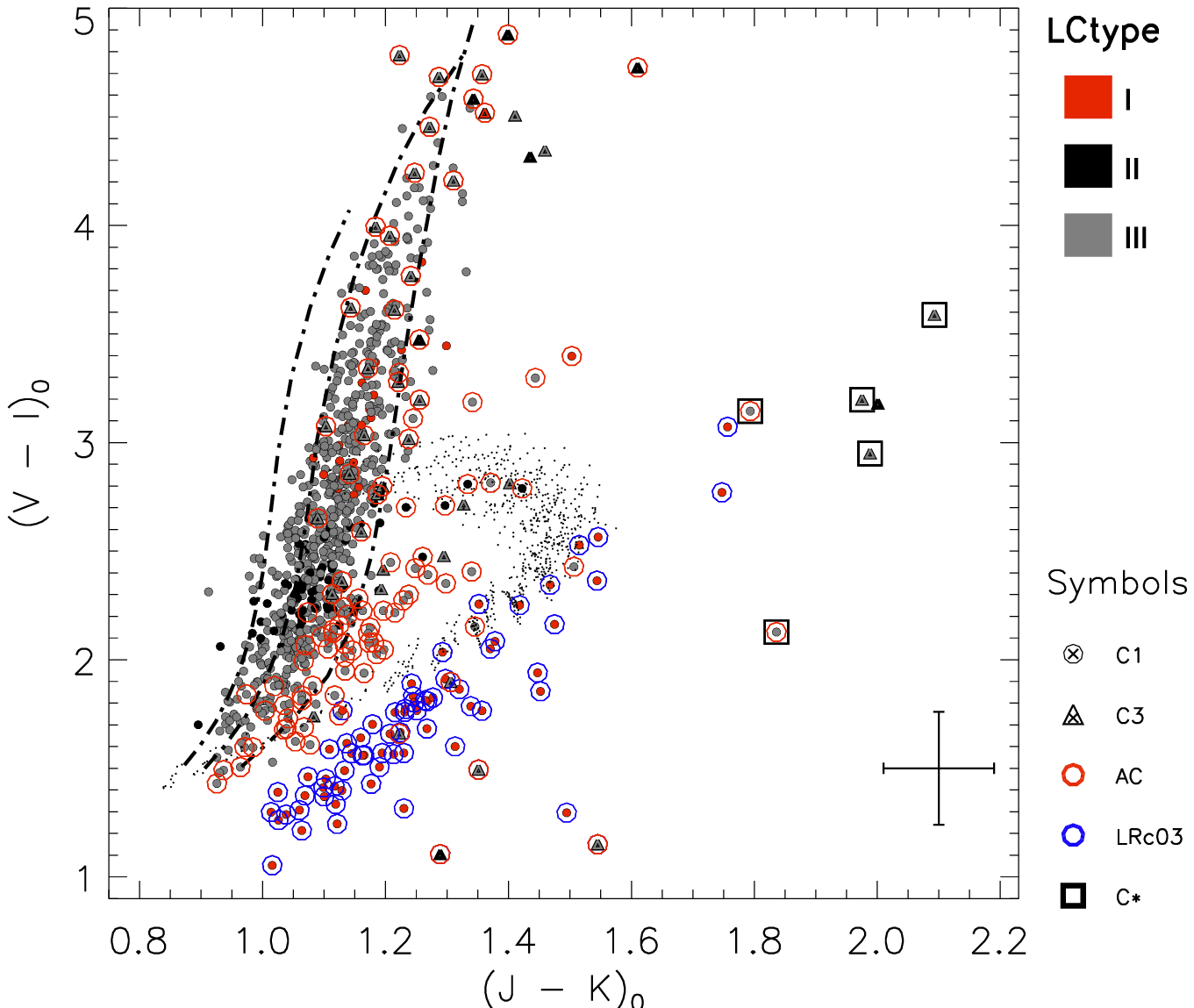}}
\caption{$V-I$ vs $J-K$ color-color diagram of C1 (circles) and C3 (triangles) subsamples. The colors indicate the spectral type (upper panel) and luminosity class (bottom panel). The dashed-dotted lines display the dereddened location of M-giants according to \citet[][]{Worthey-2011}. The small black dots mark the location of hydrostatic Carbon stars from the models by \citet[][]{Aringer-2009}. The stars previously classified as Carbon stars are depicted by a square symbol. The stars from the samples towards the outer regions of the Galaxy are marked by open red circles while the stars from the LRc03 CoRoT Run are identified using blue open circles. The error bars in the bottom right corner of the two diagrams represent the typical uncertainties for the colors ($\epsilon_{(V-I)} = 0.29$ mag and $\epsilon_{(J-K)} = 0.09$ mag)}

\label{corcor}
\end{center}
\end{figure}

The C1 and C3 sources are used to describe the evolutionary and variability behavior of M-Giant type stars in forwards sections.

%
\section{Results}\label{results}

We report our results in the following subsections, using a final sample of $1314$ M-giants (C1 and C3 stars). First, we perform an analysis of the V-I versus J-K color diagram, and compare our results to the models of \citet{Aringer-2009} and the empirical calibrations of \citet{Worthey-2011}, and to other works from the literature. Then we discuss and interpret our results using the period-amplitude and the period-$T_{\mathrm{eff}}$ diagrams. Afterwards we compare the results of the measured period and amplitude from the stars located towards the inner and outer regions of the Galaxy. Finally, we discuss the lack of short-time variations in M giants, and make a brief mention on the Mira Variable stars and the Carbon stars we have in common with other works.

\subsection{The evolutionary behavior along the color-color diagram}\label{sec_corcor}

Figure~\ref{corcor} shows the color-color diagrams of our C1 (full circles) and C3 ($T_{\mathrm{eff}} < 4200$ K - 105 stars) (full triangles) sub-samples where the colors set the spectral type (upper panel) and luminosity class (lower panel). Here we only consider stars with V, I, J and K-band photometry. The error bars in the bottom right corner of the two diagrams represent the typical uncertainties for the colors ($\epsilon_{(V-I)} = 0.29$ dex and $\epsilon_{(J-K)} = 0.09$ dex). The red and blue circles around some targets represent stars belonging to the anticenter fields and the center field LRc03 respectively. The black squares depict previously identified carbon stars, taken from the literature. The photometric calibrations of \citet[][]{Worthey-2011} are depicted by three dashed-dotted black lines, corresponding to iso-gravity contours, from left to right, of $\log g$ = 3, 2 and 0 dex. The small black dots represent the hydrostatic models of carbon giants from \citet{Aringer-2009}. We note that the reddening may not only be caused by interstellar but also by circumstellar reddening \citep[][]{Lebzelter-2011}.	

\begin{figure*}[htb]
\begin{center}
\includegraphics[width=8.0cm,height=7cm]{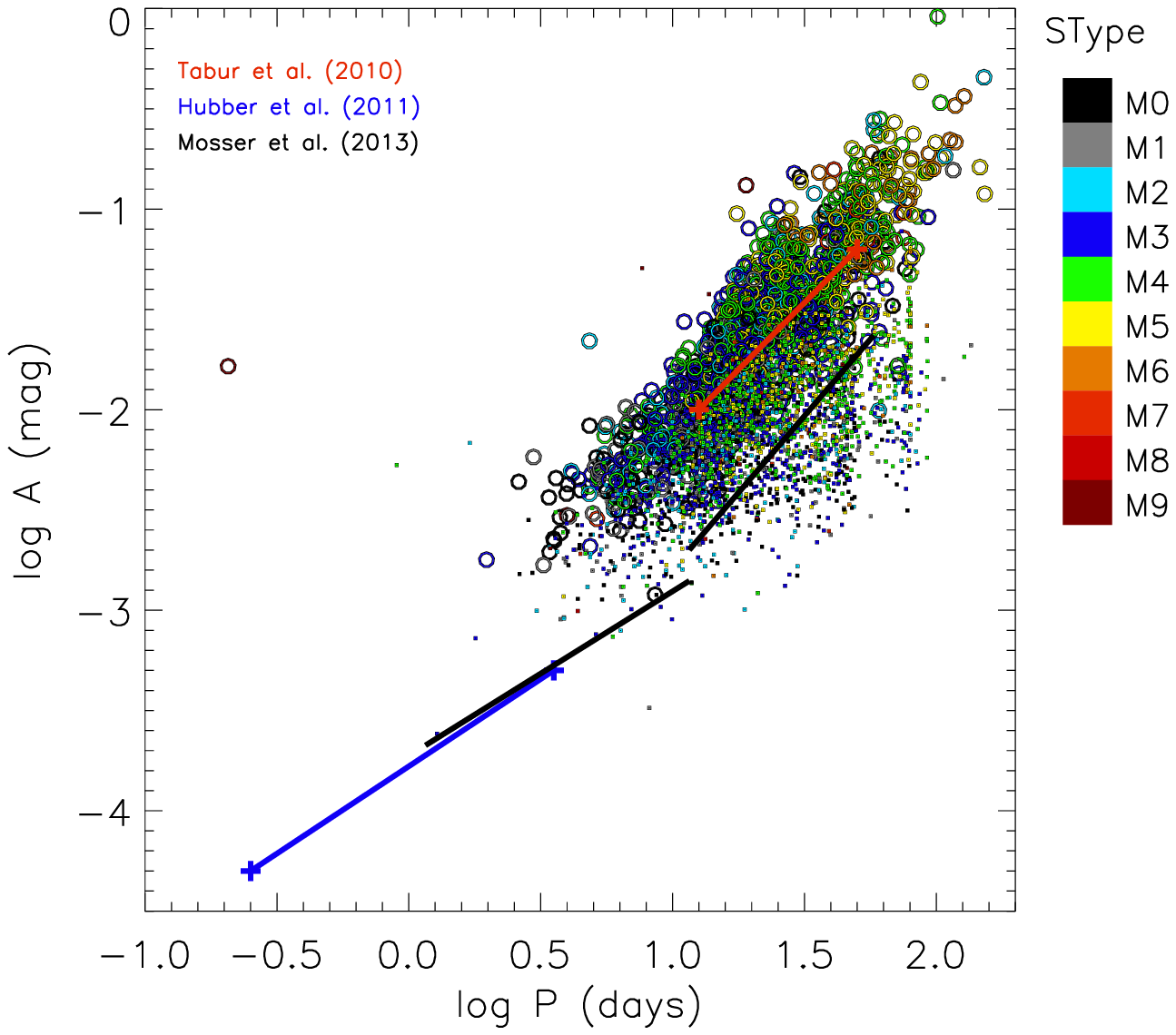}
\includegraphics[width=8.0cm,height=7cm]{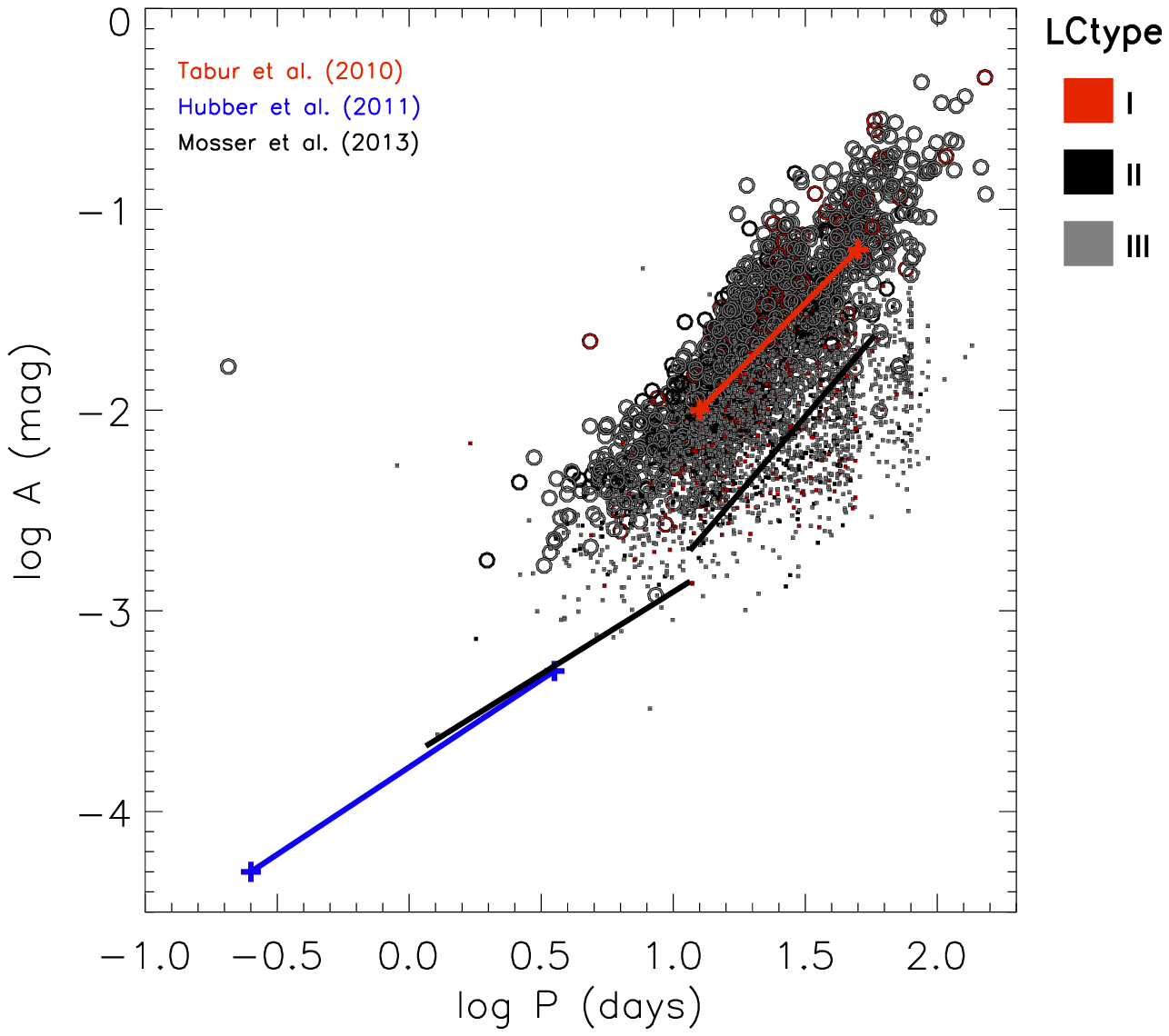}
\includegraphics[width=8.0cm,height=7cm]{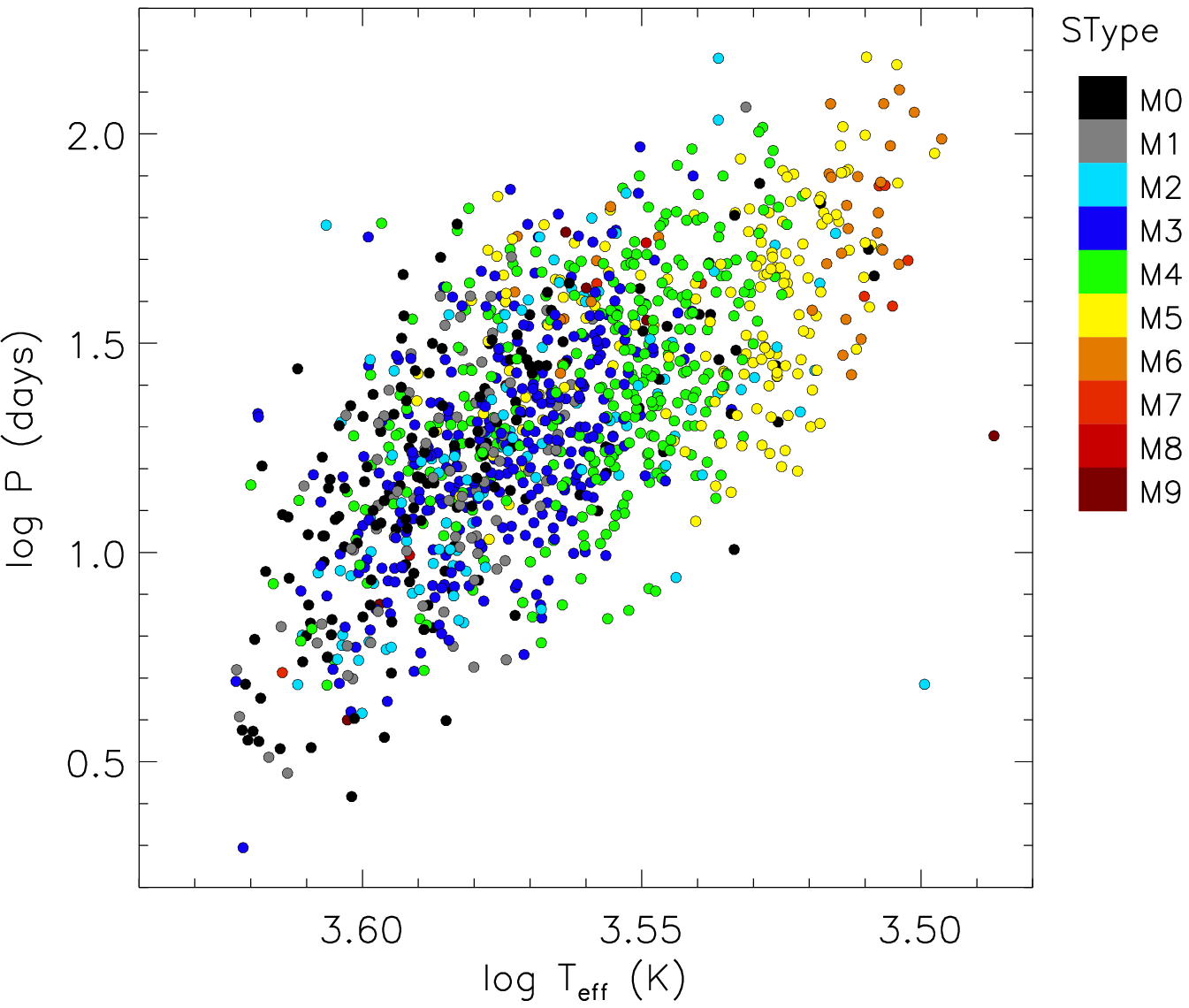}
\includegraphics[width=8.0cm,height=7cm]{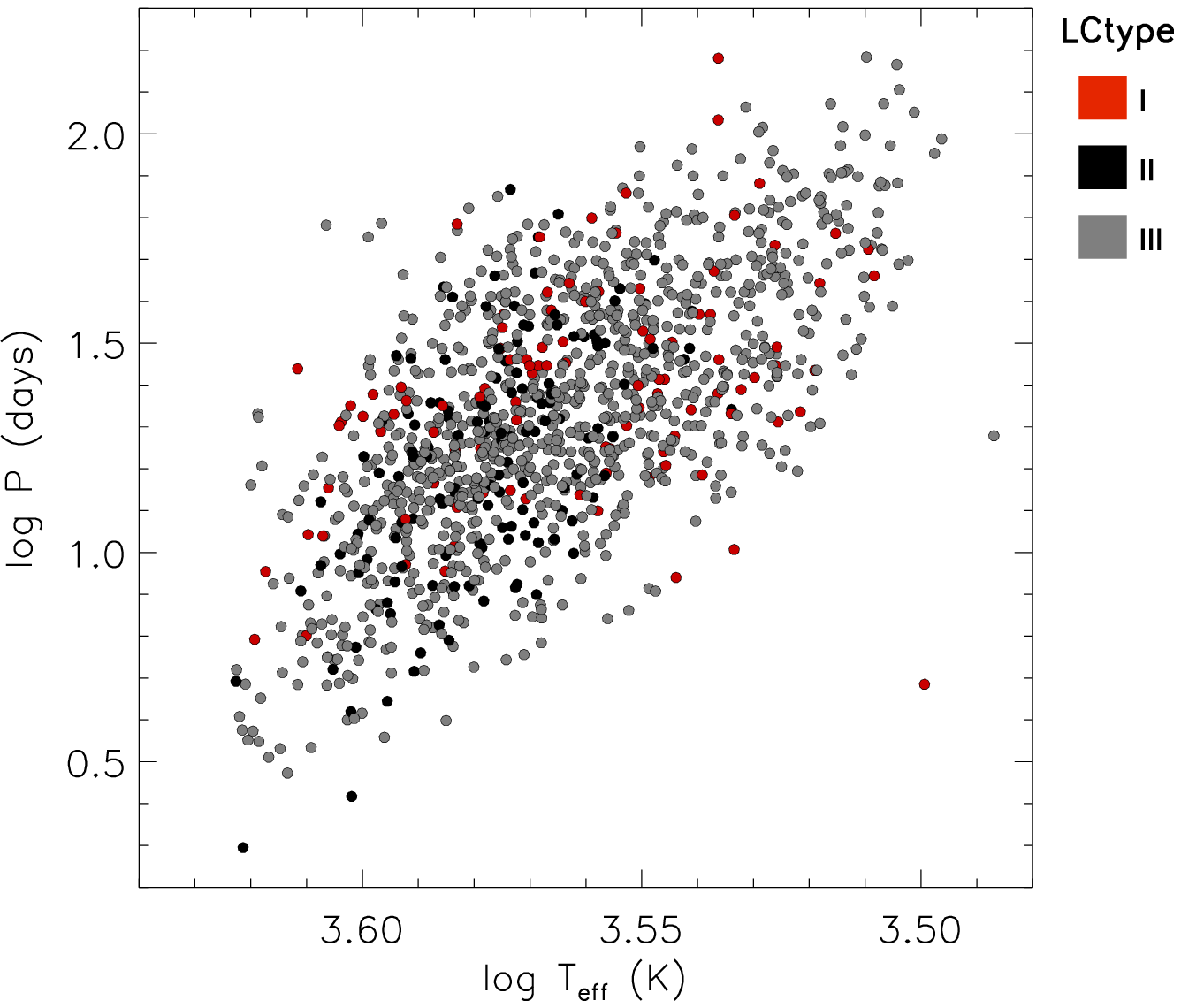}
\caption{The distribution of variability period as a function of the amplitude and the effective temperature for C1 stars. The colors indicate the spectral type (left panels) and luminosity class (right panels). The large circles in the upper panels mark the main variability period and the small circles mark the secondary periods. In the lower panels only the main variability period is shown. The red line between the two red crosses roughly depict the trend of the overtone (AB) pulsators taken from Fig. 15 of \citet[][]{Tabur-2010}. The blue line between the two blue crosses represent the area defined by \citet[][]{Banyai-2013}, selected from the top panel of Fig. 3 of \citet[][]{Huber-2011}, where is expected to find solar-like oscillations. The two black lines describe the two distinct Amplitude-$\nu_{max}$ relations for solar-type oscillators found by \citet[][]{Mosser-2013} (their Equations 8 and 9).}

\label{peramp}
\end{center}
\end{figure*}

The corrected colors show a reasonable agreement with the empirical calibrations of \citet[][]{Worthey-2011}. However, there are two groups of stars that do not follow this trend. One group, with a lower average value of $V-I$ situated between the photometric calibrations and the synthetic photometry from hydrostatic models of C-rich giants from \citet[][]{Aringer-2009}. Nevertheless, only an abundance analysis would allow us to determine if these targets are Carbon stars. Interestingly, almost all stars from this sub-group belong to the CoRoT anticenter fields. We do not know the reason of this behaviour, but we hypothesise that the samples towards the inner and outer regions of the Galaxy may belong to two different Galactic populations. In fact, an increasing trend of the C star/M star ratio with [Fe/H] in the Galaxy is reported in the literature \citep[e.g][]{Frebel-2006} and in the Large Magellanic Cloud \citep[][]{Blanco-1980, Cioni-2003}. At the same time, a negative trend with [Fe/H] has been reported by different studies \citep[e.g][]{Boeche-2013,Bergemann-2014}, at low values of Galactic vertical height ($|Z| < 300-400$ pc). Therefore, a possible explanation for the imbalance of C stars between the two fields may be related to the negative [Fe/H] trend as a function of the Galactic Radius. Our results are in close agreement with those presented in Fig. 2 of \citet[][]{Lebzelter-2011}.

We also observe that most of the super-giant population (Type I) has a lower value of $V - I$ for the same $J - K$ color, when compared to the two groups of Giant stars, and are located below the predicted region for C-rich stars. For $J-K > 1.3$ the location of these stars coincide with the models, but theoretical traces from \citet[][]{Aringer-2009} do not cover spectral class I.  

The LPVs Candidates (C3 catalog) are distributed throughout the color-color diagram. This is expected since the time span of the LPVs range from $24$ to $145$ day, implying a greater variability period. Eight of twelve stars with $V - I > 3.8$ have a total time span greater than 100 day (and therefore a greater period). These aspects agree with the period versus Teff trend (Fig.~\ref{peramp} lower panels) since we expect long periods for stars with lower temperatures (and higher $V - I$ color). This result strongly suggests that long-term trends found for all LPVs candidates may be real.

%
\subsection{The period variability and amplitude behavior}\label{ampsec}

To investigate the nature of the stellar photometric variability identified in the present work, we studied the period variability as a function of amplitude and effective temperature for stars of the present sample that exhibit semi--sinusoidal behavior in their observed LCs. For such a purpose, we have used the periods and amplitudes for the sample of $1173$ stars described in Sect. \ref{sec_cat} (the C1 catalog). To study the multi-mode behavior we considered the first three independent frequencies on the period-amplitude diagrams of Fig. \ref{peramp}. In the case of the $T_{\mathrm{eff}}$-period diagrams, we only used the main period.

The distribution of the multi-modes as a function of the amplitude is displayed in Fig.~\ref{peramp} (upper panels). The circles represent the stronger mode, while the dots depict the other modes of oscillation. The red line between the two red crosses roughly depict the trend of the overtone (AB) pulsators taken from Fig. 15 of \citet[][]{Tabur-2010}. The blue line between the two blue crosses represent the area defined by \citet[][]{Banyai-2013}, selected from the top panel of Fig. 3 of \citet[][]{Huber-2011}, where is expected to find solar-like oscillations. The two black lines describe the two distinct Amplitude-$\nu_{max}$ relations for solar-type oscillators found by \citet[][]{Mosser-2013} (their Equations 8 and 9).

The observed period-amplitude behavior clearly parallels the classical scenario for the pulsation period-amplitude relation which has been well established for M-giant stars with semi-sinusoidal variations using different stellar samples observed from the ground \citep[e.g.,][]{Alard-2001,Wray-2004,Tabur-2010}. 

The scenario presented in Fig.~\ref{peramp} (upper panels), for the stronger pulsation (open circles), follows closely the trend observed by \citet[][]{Tabur-2010} (red line in the Diagram), which corresponds to stars pulsating in overtones. The other modes of oscillation agree well with the Amplitude-$\nu_{max}$ relation of \citet[][]{Mosser-2013}, for $\nu < 1 \mu$Hz, corresponding to $\log P > 1.06$ day. The two distinct relations suggest the presence of two different oscillation mechanisms: the stronger one, characteristic of self-excited Mira-like pulsations; the smaller amplitude ones, mimicking the behavior of solar-type oscillations.

Our finding also agrees with the results of \citet[][]{Banyai-2013},  who, on the basis of \emph{Kepler} observations of M-giant stars (their group 3), found a similar period-amplitude behavior. Group 3, as defined by those authors, contains stars with LCs that exhibit only a few periodic components (like Miras and SRs) that are in agreement with the signatures of semi-sinusoidal variables. \citet[][]{Banyai-2013} also reported a sharp break at $\log P$  $\sim$ 1 day (corresponding to 1.2 $\mu$Hz), marking the end of the clear correlation between their Group 2 and their adapted relation from Amplitude-$\nu_{max}$ relation of \citet[][]{Huber-2011}. We do not observe a trend in our Fig. 3 (upper panels) in the same region of $\log P$. However, we must note we have very few periods with similar amplitudes in that Period region. Despite that, if we account only the stronger pulsation we observe, by eye, that there is a slight change of slope around $\log P \sim$ 1 day. 

We also analyzed the relation period-$T_{\mathrm{eff}}$. Fig.~\ref{peramp} (lower panels) shows all $1141$ giant stars from our C1 sample with semi-sinusoidal variations, where only the primary computed pulsation period is considered. We observe a correlation of the period of the main oscillation with $T_{\mathrm{eff}}$ similar to the trend found by \citet[][]{Huber-2011} for hotter giants from the \emph{Kepler} sample,  with the oscillation frequency $v_{max}$. A careful comparison of our Fig.~\ref{peramp} (bottom left panel)  with Fig. 8 of \citet[][]{Huber-2011}, points for an agreement between both trends. Indeed, the feature revealed from Fig.~\ref{peramp} (lower panels) seems to correspond to an extension of the upper part of the modified H--R diagram (Figure 8) of \citet[][]{Huber-2011}.

\begin{figure}[htb]
\begin{center}
\includegraphics[height=7cm]{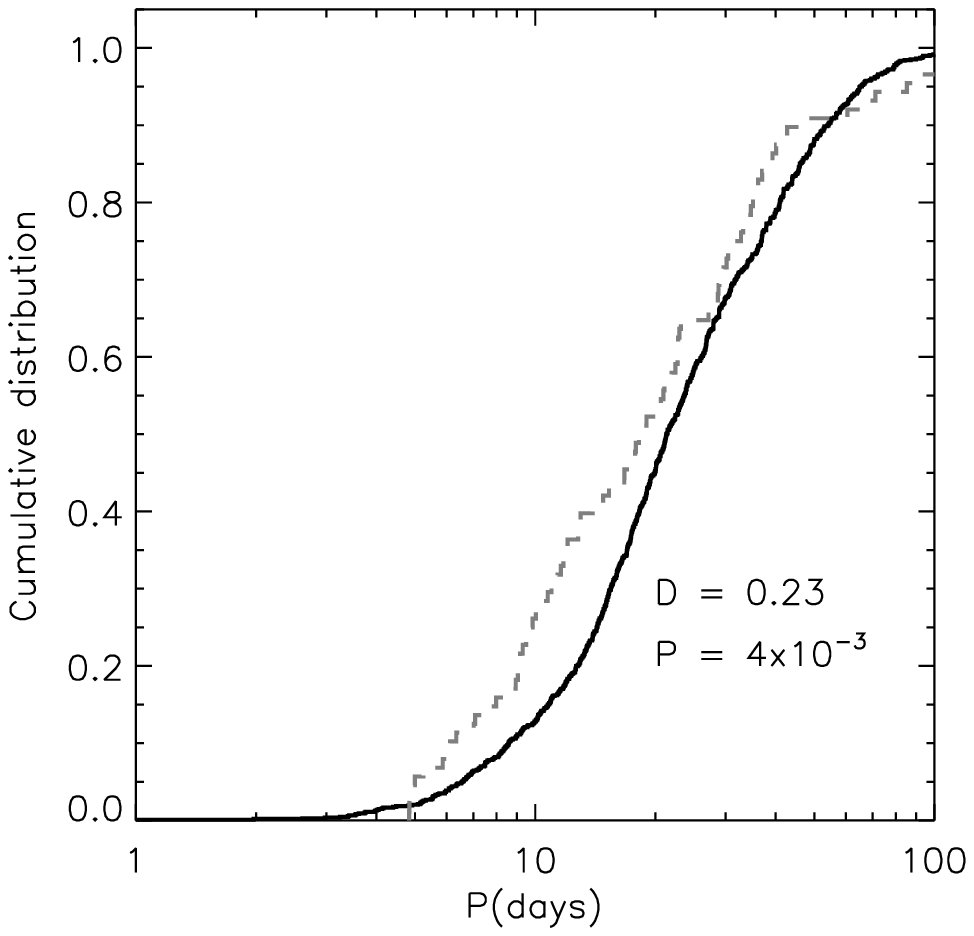}
\includegraphics[height=7cm]{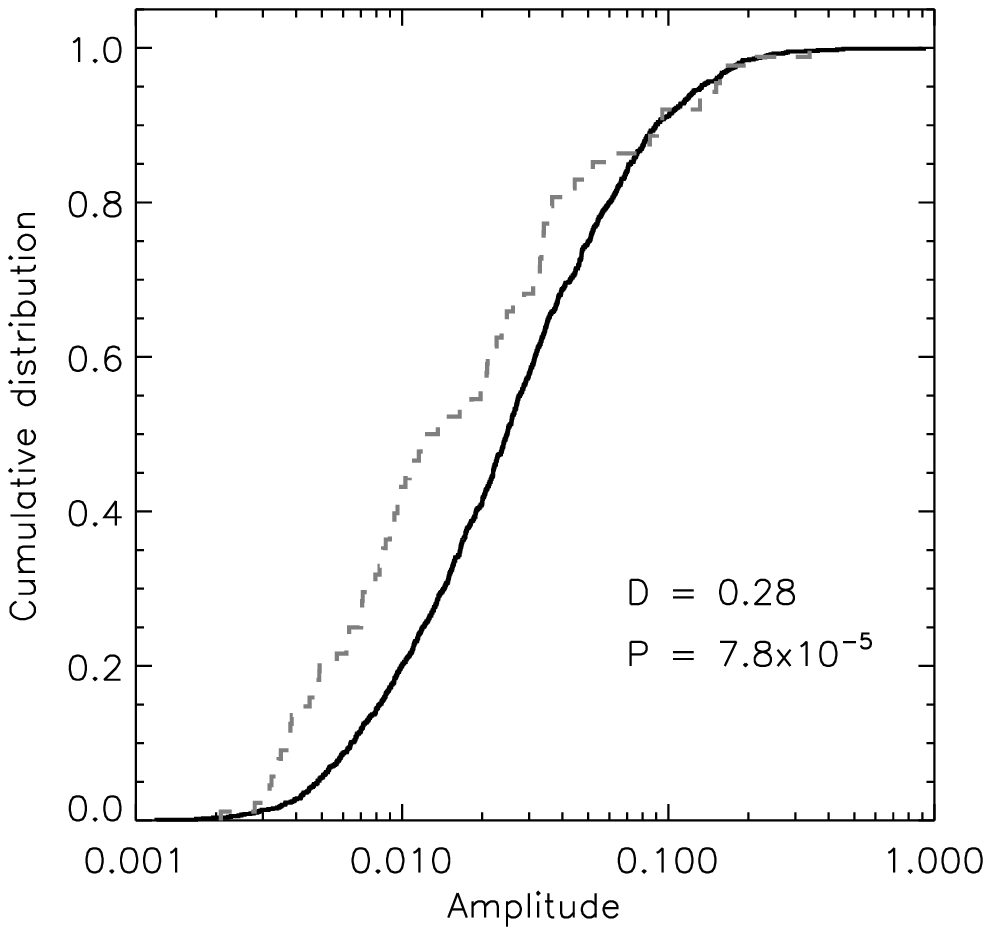}
\caption{Cumulative distribution function of the period (left panel) and amplitude (right panel) for stars at the inner (solid line) and outer (dashed line) regions of the Galaxy.}

\label{kstest}
\end{center}
\end{figure}

\begin{figure*}[tb]
\begin{center}
\includegraphics[width=8.0cm,height=6cm]{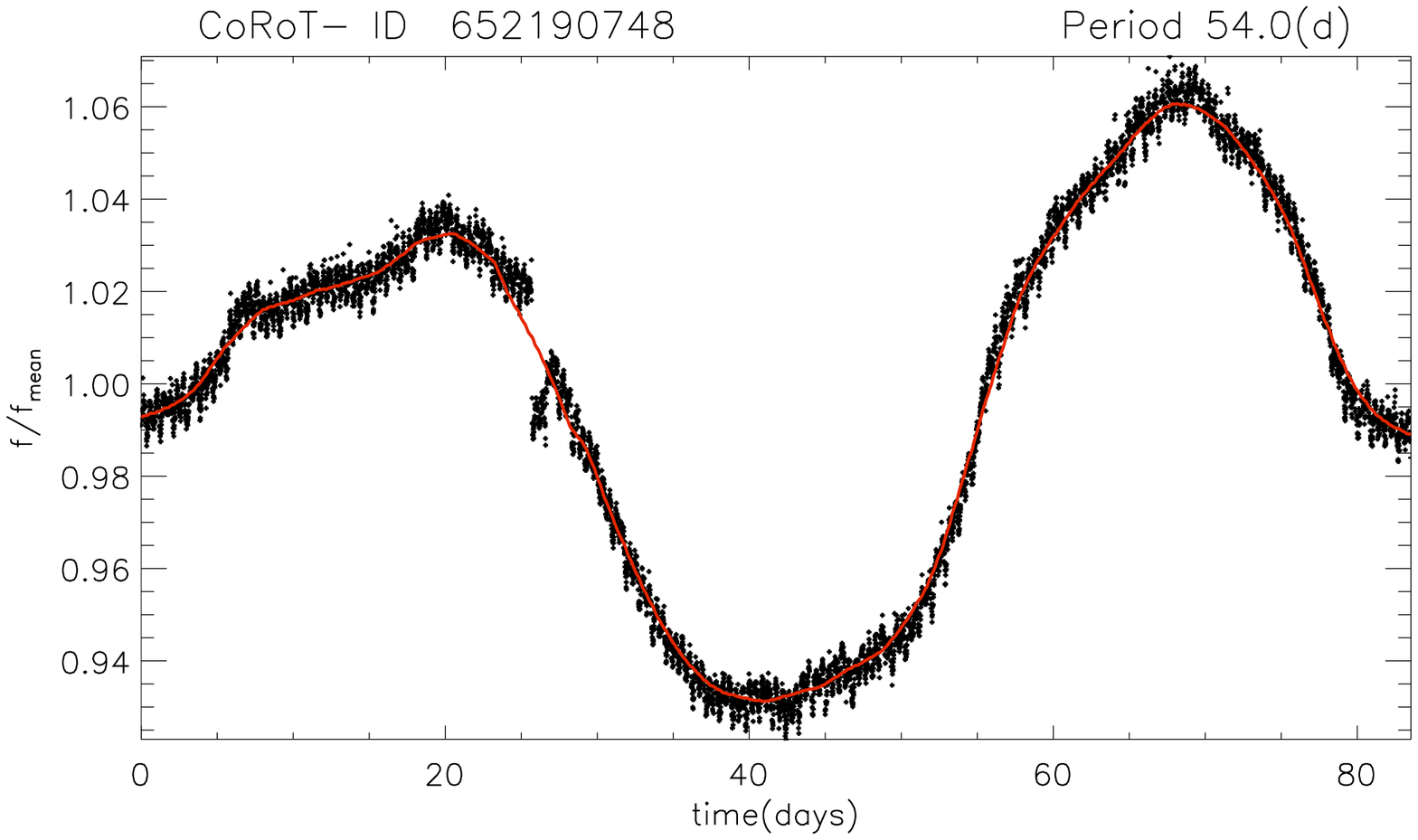}
\includegraphics[width=8.0cm,height=6cm]{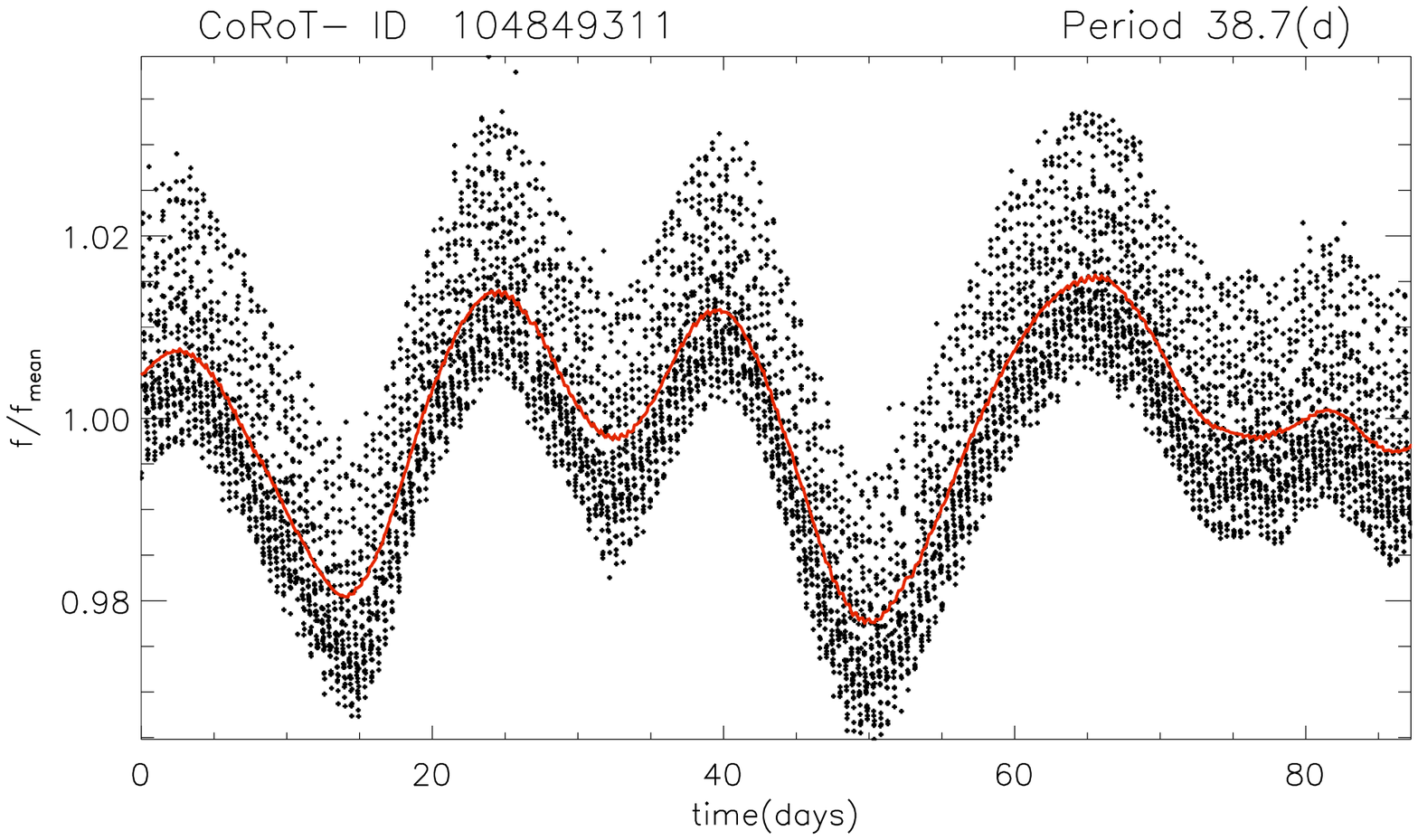}

\includegraphics[width=8.0cm,height=4cm]{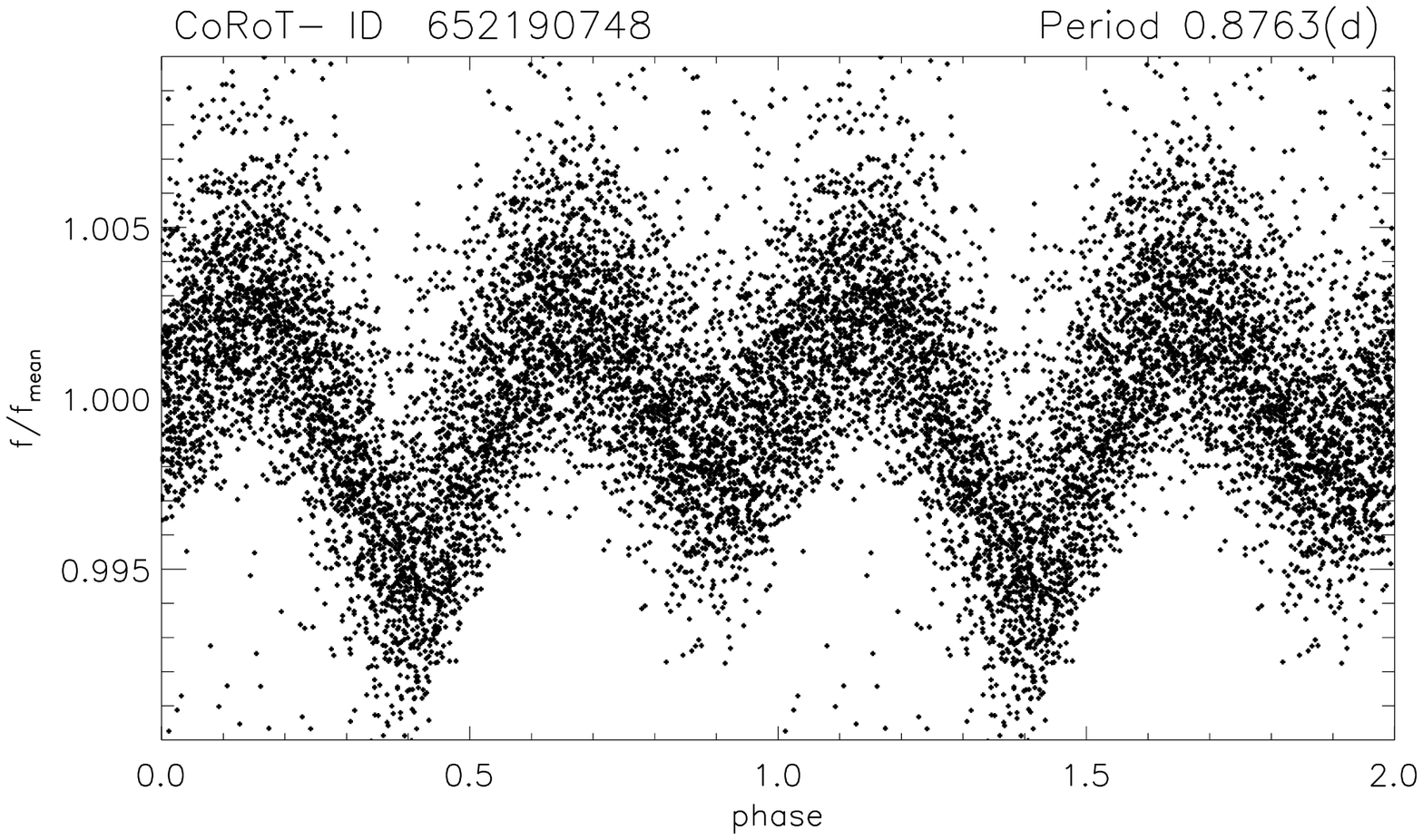}
\includegraphics[width=8.0cm,height=4cm]{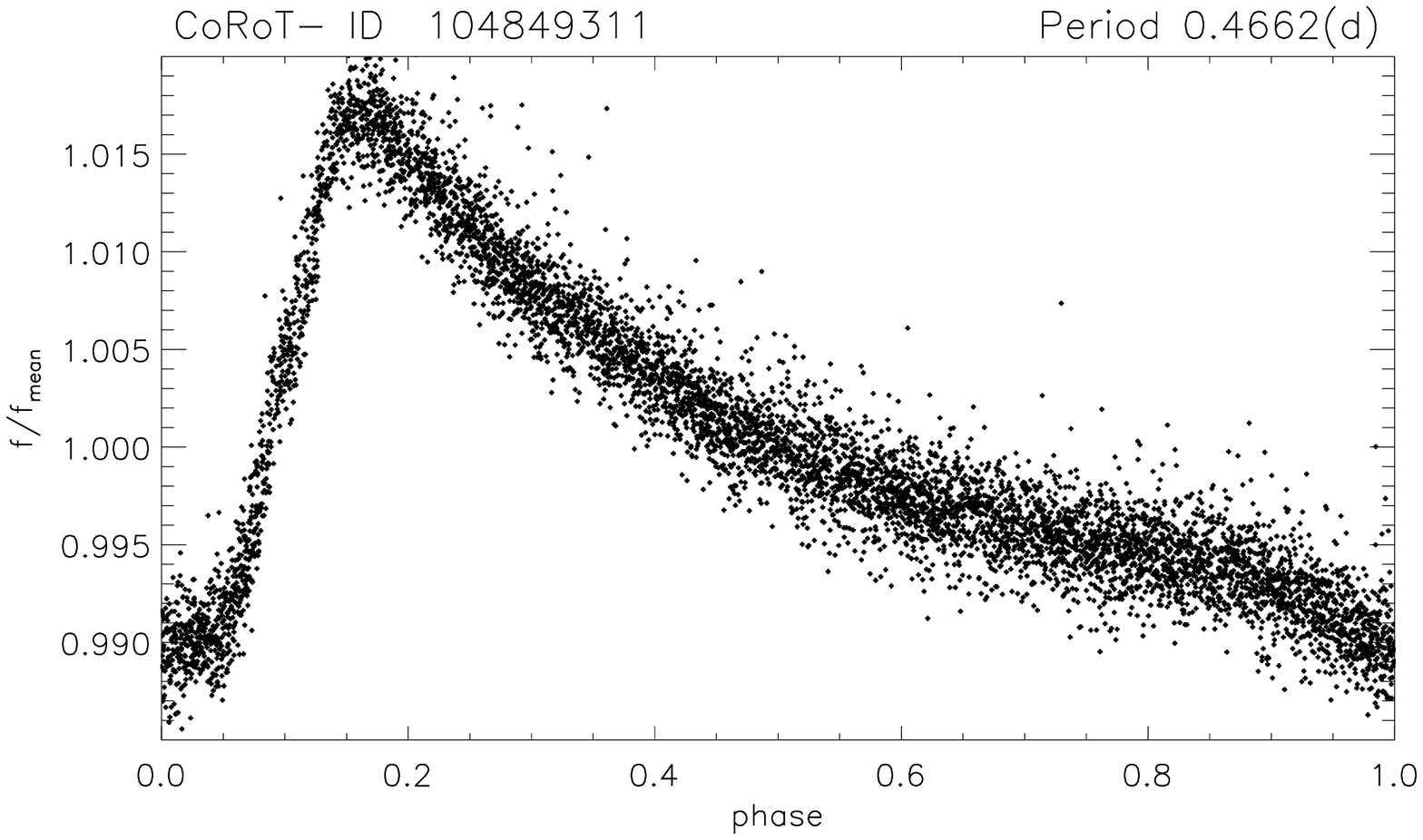}

\caption{CoRoT 652190748 and CoRoT 104849311 light curves (upper panels). The red line displays a boxcar smoothed version of the LC. The phase diagrams of their short time variations are depicted in the lower panels. The CoRoT-ID and period are presented in each panel.}
\label{figshort}
\end{center}
\end{figure*} 

\subsection{The inner and outer regions of the Galaxy}\label{kstest}

To verify whether the behavior of the variability of sources from the inner and outer regions of the Galaxy are significantly different, we applied the Kuiper test (or invariant KS test - hereafter the KS test) \citep[e.g. ][]{Jetsu-1996,Paltani-2004} to our C1 subgroup. The C1 is composed of $1124$ stars located toward inner and $91$ in outer regions of the Galaxy. According to the KS test, zero probability means that the distributions are dissimilar, whereas unit probability means they are drawn from the same parent distribution.

Figure \ref{kstest} shows the cumulative distribution function of the amplitude and period of C1 stars. The solid black distribution depicts the stars located in the inner regions whereas the dashed grey distribution describes the outer region. The obtained p-values for the amplitude and period distributions are lower than $10^{-3}$. This result shows that there is a very low probability that the two samples belong to the same parent distribution. The difference between the two samples may also be observed in the color-color diagram (Fig.~\ref{corcor}) where most stars from the outer region sample are in the region commonly associated with Carbon stars, while the sample from the inner region of the Galaxy are mostly M giants. Also, the metallicity of our two samples should be different (see Sect. \ref{sec_corcor}) and this may play an important hole on M Giant evolution, since these distributions are dissimilar in terms of amplitude and period.

We tested the KS test by perturbing our two samples. First, we withdrew a random quantity of stars from our sample between $5\%$ to $10\%$ using an uniform Monte Carlo distribution and next we performed the KS test. The process was iterated for 1000 times. The higher $3\sigma$ value taken from the median of the amplitude and period distribution was $0.002$ and $0.05$ respectively. The indicative of dissimilarity is not robust enough for the period distribution.

%
\subsection{Short-time vartiations of CoRoT M-giant stars}\label{shortvar}

The CoRoT data have at least a 512 second sampling interval and a long coverage period that allow us to study the occurrence of rapid outbursts or very short time variations in LPVs on very short time scales (from hours to a few days). These variation changes of several tenths of magnitudes should occur once per star per year according to the literature \citep[][]{Wozniak-2004}. A recent study using a sample of 52 CoRoT stars \citep[][]{Lebzelter-2011} and 9 \emph{Kepler} stars \citep[][]{Hartig-2014}  presented no detections of such rapid variations for amplitudes above 0.01 mag. However, only a small number of sources was analysed.

A detailed search by visual inspection on the LCs of all CoRoT M giant stars composing the present sample, did not yield any short term event. All outburst-like variation found may be related to hot pixel events or other instrumental artefacts (check, for instance, the LC panels of CoRoT 105731070, 105832319, and 105965810 of Fig.~\ref{lc_c1}). This agrees with other recent studies showing that this phenomenon is rare or does not exist \citep[e.g.,][]{Wozniak-2004,Lebzelter-2011,Hartig-2014}. Taking into account all results from different instruments and small and large samples, we can conclude that the short-time outbursts (< 1 day) in long-period variables do not exist or at least are very rare events.

Another possible cause of a false outburst event may lie with a background star. In this case, the observed variation will be the superposition of the signal from both stars. Fig.~\ref{figshort} shows two CoRoT LPVs where the phase diagrams (lower panels) show a signature of a background binary (left panel) and a RR-Lyrae (right panel) star. To study the short time variation of the two LCs, a prewhitened curve was obtained by dividing each LC by a boxcar smoothing function of the curve itself. Then, the period was computed using a similar proceedure  such as described in Sect.~\ref{submeth}.

%
\subsection{Long Period Variables}\label{ssmira}

Three LPVs stars of our sample were previously analyzed by \citet[][]{Lebzelter-2011} using CoRoT data. Some of these stars are also common to the AAVSO International Variable Star Index \citep[][]{Watson-2014} catalog. In the present LPVs sample (C3) we identify four Mira variable stars,  CoRoT 104982243, 105580931, 632685506, and 653546843, with previously computed periods of 329, 205, 195, and 116 day respectively \citep[][]{Watson-2014}.

%
\subsection{Carbon Type Stars}\label{sscarb}

Eight carbon type stars, previously described in the literature \citep[][]{MacConnell-1988,Stephenson-1989,Stephenson-1996,Alksnis-2001,Chen-2012,Zacharias-2013,Watson-2014}, are identified in our sample. These stars are in the fields towards the outer regions of the Galaxy (LRa01, IRa01, and LRa02 CoRoT runs). The combination of different CoRoT Runs allow us to study the long periods of the Carbon stars with more accuracy. For instance, CoRoT-ID 102760221 (see Fig.~\ref{lc_c1} last row, right panel) was analysed by Lebzelter (2011) using the CoRoT Runs IRa01 and LRa01. A period of 116 days is reported by \citet{Lebzelter-2011} as the probable main period once its LC present a minimum and a maximum between 250 and 400 days. Because the referred star was re-observed during the LRa06 Run, we have combined all the observations (CoRoT Runs IRa01, LRa01 and LRa06) for a check of the period measurements, finding now a period of 349 days, which may be related with the fundamental pulsation mode.

\begin{figure}[tb]
\begin{center}
\includegraphics[width=8.0cm,height=5cm]{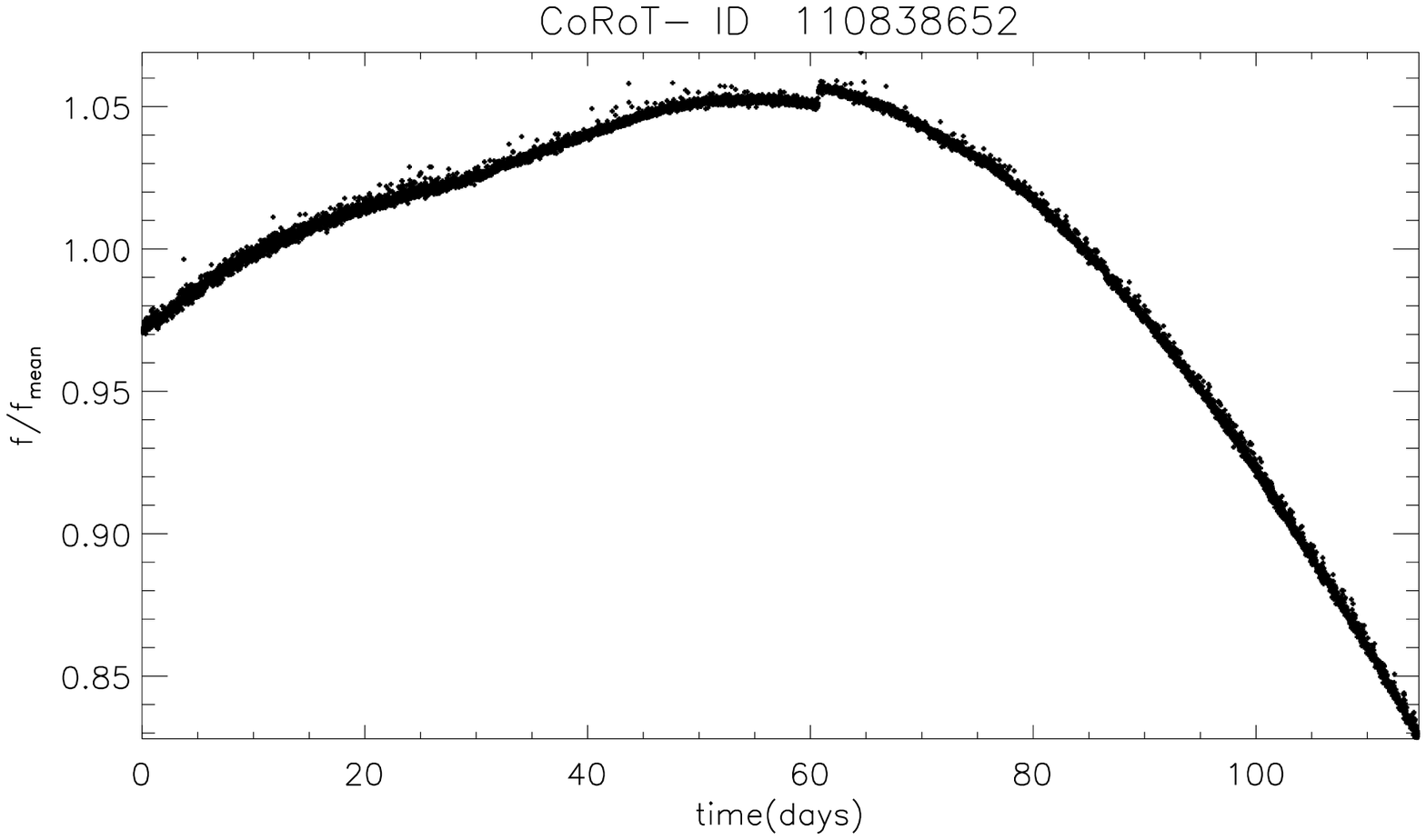}
\includegraphics[width=8.0cm,height=5cm]{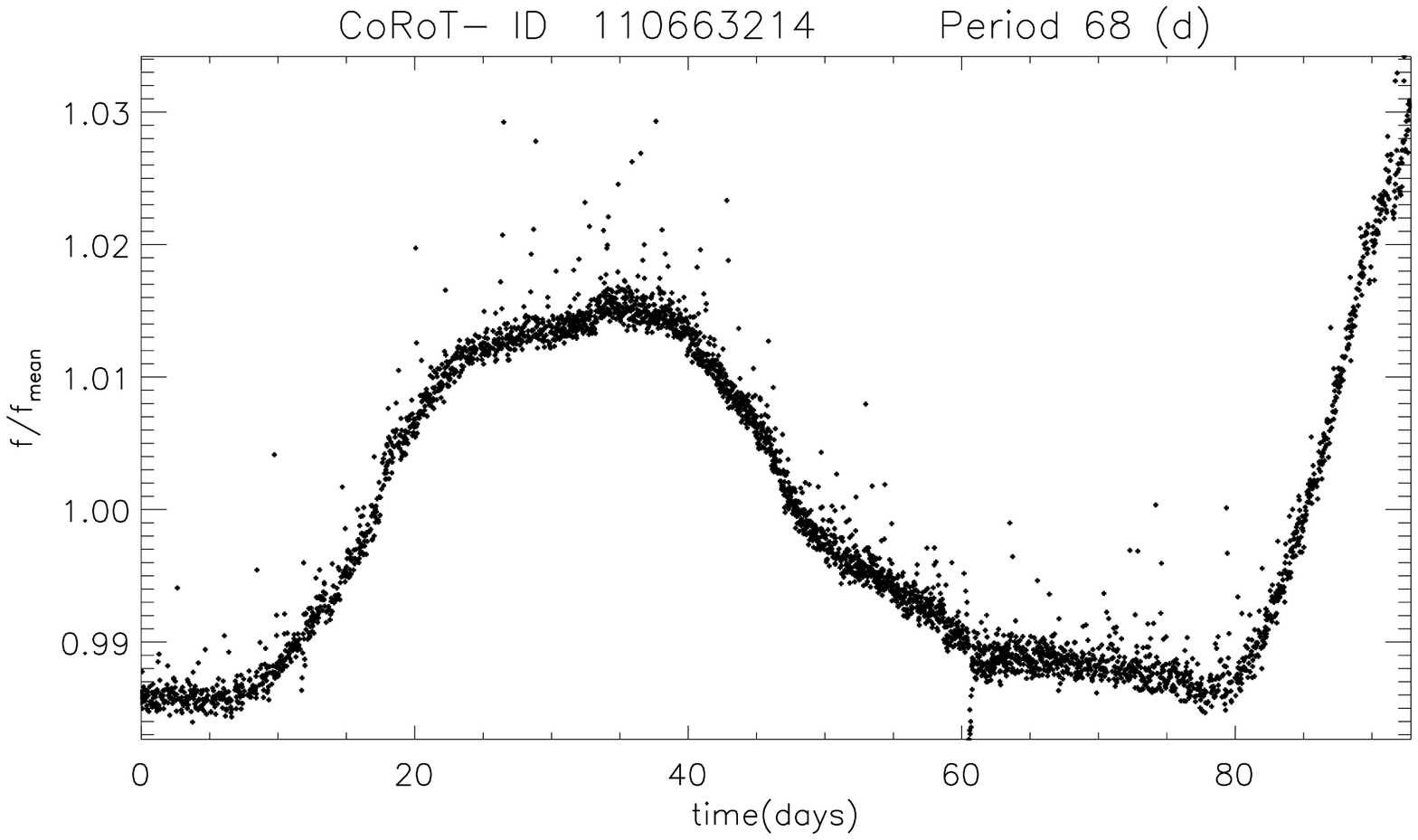}

\caption{LCs of two previously identified carbon type stars: CoRoT 110838652 and 110663214.}

\label{figcarb}
\end{center}
\end{figure}

On the other hand, in some cases, the total time span is not long enough to provide a reliable period (see Fig.~\ref{figcarb}, upper panel) while in other cases we can observe more than one cycle (Fig.~\ref{figcarb}, lower panel). For instance, the star CoRoT 110663214 displays a period of 68 day. However, the shape of its LC indicates the existence of a larger period. Using the methodology of \citet{De-Medeiros-2013} we  obtain a confidence level of about $70\%$ for this period.

%
\section{Conclusions}\label{conclusions}

In this work we perform a study of the semi-sinusoidal variability behavior of CoRoT M-giant stars. This is the largest detailed study of M-giants observed by CoRoT, and it provides an opportunity for comparison with previous works based on space missions like CoRoT and \emph{Kepler} \citep[e.g.][]{Lebzelter-2011,Banyai-2013, Huber-2014,Hartig-2014} and on ground-based observations \citep[e.g.,][]{Alard-2001, Wray-2004,Tabur-2010}.

We also present a CoRoT variability list for M--giant stars. The C1 sample, which contains $1173$ objects, is composed of all stars classified as M-giants in the CoRoT database with $T_{\mathrm{eff}} < 4200$~K. They were identified via visual inspection as semi-sinusoidal variables. The effective temperature $T_{\mathrm{eff}}$ was estimated from the calibration of \citet[][]{van-Belle-1999} using $V-K$ color indices. The main variability period spans from $\sim 2$ to $\sim 152$ day, with the amplitude ranging from $\sim 1$ to $\sim 900$~mmag.  The C2 sample is comprised of $141$ stars that exhibit semi-sinusoidal variations (with $T_{\mathrm{eff}} > 4200$K), and the C3 sample contains $114$ LPV candidates. The two last sub-samples may be considered for follow-up observations to study their variability nature.

We performed a cross-check of our data with previously published catalogs and found 444 targets in common. The computed periods are similar (i.e., discrepancy less than $10\%$) for $96\%$ of these targets, which indicates a good agreement between our results and previous results from the literature. The cross-matched sources were used to withdraw misclassification and to identify comparison stars.

The location of the majority of the sample stars in the $V-I$ versus $J-K$ color-color diagram shows a good agreement with the empirical calibrations of \citet{Worthey-2011}. Two groups of stars present different trends. The first one, with a lower average value of $V-I$ for the same $J-K$ is mostly composed with stars from the outer region sample and is located between the Giant calibrations and the Carbon star models of \citet{Aringer-2009}. The second group, located in the lowest $V-I$ region of the diagram, appears to be mostly composed by a Type I supergiant population. The large majority of the stars from this group belong to the LRc03 field, towards the inner regions of the Galaxy. 

We considered the first three periods with significance levels greater than $99\%$ to study their behavior in a period-amplitude diagram. The distribution exhibits a trend of increasing amplitude with increasing period, which is compatible with the expected behavior for stellar pulsation  \citep[e.g.,][]{Alard-2001,Wray-2004,Tabur-2010}. We also observe that the main variability period follows closely the stars identified by \citet{Tabur-2010} as overtone pulsators. The less powerful periods follow the trend formulated by \citet{Mosser-2013} for solar-like oscillations. A similar analysis was performed using a period versus $T_{\mathrm{eff}}$ diagram, which exhibits a trend of increasing period with decreasing temperature. This trend is compatible with a previous study based on \emph{Kepler} observations for GK giants \citep[][]{Huber-2011}, and our new data confirms that the behavior of the period--effective temperature relation extends to the cooler M-giant stars.

Regarding the short-time variations on M-giant stars previously reported in the literature, we also concluded that they do not exist or are very rare. The few cases we found are related to biases or background stars.

\section*{Acknowledgments}

We would like to thank Evelin Banyai for useful discussions. Research activities of the Observational Stellar Board of the Federal University of Rio Grande do Norte are supported by the continuous grants of the CNPq and FAPERN Brazilian agencies and by the INCT-INEspa\c{c}o. C. E. F. L. acknowledges a post-doctoral fellowship from the CNPq. F. P. Ch. acknowledges a doctoral fellowship from the CNPq. This research has made use of the NASA/IPAC Infrared Science Archive, which is operated by the Jet Propulsion Laboratory, California Institute of Technology, under contract with the National Aeronautics and Space Administration.

\bibliographystyle{aa}
\bibliography{mylib02}

\begin{table*}[h!]
\caption{Stellar parameters of C1.}
\label{tab-cat01}
\centering
\scalebox{0.55}{
\begin{tabular}{l c c c c c c c c c c c c c c c c c c c c}
\hline
\hline
ID(CoRoT) & ID & RA[deg.] & DEC[deg.] & ST & LC &  Type & V & $V_{0}$ & I & $I_{0}$ & J & $J_{0}$ & K & $K_{0}$ &  T$_{eff}$(K) &  P(day)  &  A(mmag)   \\
\hline
101112162 &  UNC & 291.6292 & 1.1659 & M0 & II &  Semi-sinV* & 13.90 & 12.32$\pm$0.17 & 11.84 & 10.62$\pm$0.14 & 9.94 & 9.52$\pm$0.09 & 8.81 & 8.63$\pm$0.06 & 3999$\pm$102 & 2.61, 16.4, 4.44 & 4.37, 2.85, 2.03 \\ 
102285797 &  102285797 & 92.2382 & 4.5606 & M3 & III &  Semi-sinV* & 12.91 & 12.08$\pm$0.25 & - & - & 8.49 & 8.27$\pm$0.13 & 7.21 & 7.12$\pm$0.09 & 3591$\pm$68 & 19.0, 17.1, 56.1 & 36.6, 12.7, 12.7 \\ 
102299158 &  102299158 & 92.4169 & 5.1898 & M0 & III &  Semi-sinV* & 12.93 & 12.32$\pm$0.22 & 10.85 & 10.24$\pm$0.18 & 9.12 & 8.96$\pm$0.11 & 7.89 & 7.82$\pm$0.08 & 3716$\pm$76 & 37.0, 16.8, 25.1 & 28.7, 11.6, 13.4 \\ 
102311422 &  UNC & 92.5670 & 4.4369 & M6 & III &  Semi-sinV* & 14.50 & 13.70$\pm$0.26 & - & - & 7.64 & 7.42$\pm$0.14 & 6.20 & 6.11$\pm$0.09 & 3211$\pm$47 & 118, 79.8, 29.9 & 216, 27.8, 12.3 \\ 
102335279 &  UNC & 92.8541 & 5.1593 & M4 & III &  Semi-sinV* & 13.15 & 12.55$\pm$0.24 & 10.20 & 9.44$\pm$0.19 & 7.87 & 7.70$\pm$0.12 & 6.53 & 6.46$\pm$0.08 & 3375$\pm$51 & 104, 26.8, 43.4 & 339, 20.9, 18.5 \\ 
102338552 &  102338552 & 92.8902 & 5.0779 & M1 & III &  Semi-sinV* & 12.93 & 12.24$\pm$0.24 & 10.72 & 10.03$\pm$0.19 & 8.92 & 8.74$\pm$0.12 & 7.68 & 7.60$\pm$0.08 & 3674$\pm$73 & 22.4, 32.2, 27.4 & 24.7, 20.7, 13.1 \\ 
102349003 &  102349003 & 93.0078 & 4.5821 & M1 & III &  Semi-sinV* & 14.25 & 13.25$\pm$0.21 & 11.92 & 11.03$\pm$0.17 & 9.95 & 9.68$\pm$0.11 & 8.63 & 8.52$\pm$0.07 & 3647$\pm$62 & 23.0, 26.0, 70.1 & 23.6, 15.4, 17.6 \\ 
102354589 &  102354589 & 93.0736 & 4.3677 & M2 & III &  Semi-sinV* & 14.06 & 12.88$\pm$0.21 & 11.68 & 10.67$\pm$0.17 & 9.69 & 9.38$\pm$0.11 & 8.30 & 8.16$\pm$0.08 & 3652$\pm$63 & 42.7, 58.3, 41.2 & 58.9, 26.0, 10.2 \\ 
102369726 &  UNC & 93.2538 & 5.1756 & M3 & II &  Semi-sinV* & 13.55 & 12.58$\pm$0.22 & 10.82 & 9.87$\pm$0.18 & 8.63 & 8.37$\pm$0.11 & 7.25 & 7.14$\pm$0.08 & 3487$\pm$55 & 28.9, 45.3, 36.2 & 151, 24.9, 14.1 \\ 
102388262 &  102388262 & 93.4664 & 4.8155 & M5 & III &  Semi-sinV* & 13.95 & 12.89$\pm$0.24 & 10.67 & 9.57$\pm$0.19 & 8.09 & 7.81$\pm$0.12 & 6.71 & 6.59$\pm$0.08 & 3344$\pm$48 & 17.5, 20.0, 24.2 & 94.9, 20.9, 19.1 \\ 
\hline
\hline
\end{tabular}
}
\end{table*}

\begin{table*}[h!]
\caption{Stellar parameters of C2.}
\label{tab-cat02}
\centering
\scalebox{0.55}{
\begin{tabular}{l c c c c c c c c c c c c c c c c c c c c}
\hline
\hline
ID(CoRoT) & ID & RA[deg.] & DEC[deg.] & ST & LC &  Type & V & $V_{0}$ & I & $I_{0}$ & J & $J_{0}$ & K & $K_{0}$ &  T$_{eff}$(K) &  P(day)  &  A(mmag)   \\
\hline
102371380 &  UNC & 93.2726 & 5.4890 & M1 & III &  Semi-sinV* & 13.59 & 12.57$\pm$0.22 & 12.06 & 11.26$\pm$0.18 & 10.51 & 10.24$\pm$0.11 & 9.51 & 9.39$\pm$0.08 & 4238$\pm$143 & 21.5 & 126 \\ 
102573436 &  UNC & 100.1715 & 0.6706 & M1 & III &  Semi-sinV* & 12.67 & 10.63$\pm$0.22 & 10.87 & 9.43$\pm$0.18 & 9.13 & 8.59$\pm$0.11 & 8.04 & 7.80$\pm$0.08 & 4434$\pm$181 & 3.94 & 19.4 \\ 
102596573 &  UNC & 100.3367 & 0.9047 & M1 & III &  Semi-sinV* & 16.00 & 13.71$\pm$0.35 & 14.11 & 12.49$\pm$0.28 & 12.31 & 11.71$\pm$0.18 & 11.28 & 11.02$\pm$0.12 & 4526$\pm$267 & 59.5 & 35.5 \\ 
102628059 &  UNC & 100.5269 & -1.3317 & M2/M2 & III/III &  Semi-sinV* & 15.92 & 14.36$\pm$0.21 & 14.09 & 12.90$\pm$0.17 & 12.49 & 12.08$\pm$0.11 & 11.52 & 11.34$\pm$0.08 & 4327$\pm$146 & 28.2 & 2.15 \\ 
102630805 &  UNC & 100.5438 & -1.1984 & M1/K3 & III/II &  Semi-sinV* & 14.45 & 13.10$\pm$0.24 & 12.87 & 11.86$\pm$0.19 & 11.33 & 10.97$\pm$0.12 & 10.39 & 10.24$\pm$0.08 & 4414$\pm$188 & 24.5 & 2.81 \\ 
102691972 &  UNC & 100.8702 & -0.4528 & M0/K4 & III/III &  Semi-sinV* & 13.25 & 12.09$\pm$0.19 & 11.65 & 10.77$\pm$0.15 & 10.18 & 9.87$\pm$0.10 & 9.12 & 8.99$\pm$0.07 & 4280$\pm$131 & 90.3 & 9.25 \\ 
102718402 &  102718402 & 101.0124 & -0.8071 & M1/M1/K3 & III/III/II &  Semi-sinV* & 15.91 & 14.37$\pm$0.23 & 14.23 & 13.10$\pm$0.18 & 12.61 & 12.20$\pm$0.11 & 11.59 & 11.41$\pm$0.08 & 4361$\pm$178 & 11.6 & 1.87 \\ 
102726142 &  UNC & 101.0598 & -1.2476 & M2 & III &  Semi-sinV* & 12.61 & 10.67$\pm$0.30 & 10.75 & 9.34$\pm$0.24 & 9.01 & 8.50$\pm$0.15 & 7.79 & 7.57$\pm$0.10 & 4279$\pm$186 & 14.7 & 4.07 \\ 
102739465 &  UNC & 101.1351 & 1.0326 & M1 & III &  Semi-sinV* & 13.38 & 11.62$\pm$0.41 & 11.59 & 10.31$\pm$0.33 & 9.91 & 9.44$\pm$0.21 & 8.79 & 8.58$\pm$0.14 & 4316$\pm$255 & 2.94 & 2.92 \\ 
102748357 &  102748357 & 101.1845 & -1.6619 & M1 & III &  Semi-sinV* & 16.05 & 13.92$\pm$0.17 & 14.00 & 12.43$\pm$0.13 & 12.54 & 11.98$\pm$0.08 & 11.60 & 11.36$\pm$0.05 & 4606$\pm$167 & 8.18 & 55.0 \\ 
\hline
\hline
\end{tabular}
}
\end{table*}

\begin{table*}[h!]
\caption{Stellar parameters of C3.}
\label{tab-cat03}
\centering
\scalebox{0.55}{
\begin{tabular}{l c c c c c c c c c c c c c c c c c c }
\hline
\hline
ID(CoRoT) & ID & RA[deg.] & DEC[deg.] & ST & LC &  Type & V & $V_{0}$ & I & $I_{0}$ & J & $J_{0}$ & K & $K_{0}$ &  T$_{eff}$(K)  \\
\hline
102659593 &   IRAS 06401+0057 & 100.7014 & 0.9100 & M1 & III &  LPV* & 15.35 & 13.69$\pm$0.25 & 11.91 & 10.51$\pm$0.20 & 8.98 & 8.54$\pm$0.13 & 6.73 & 6.54$\pm$0.09 & 3248$\pm$75 \\ 
102753101 &  DJM  35 & 101.2107 & 0.8374 & M1 & III & C* & 15.73 & 14.05$\pm$0.28 & 11.59 & 10.13$\pm$0.22 & 9.25 & 8.80$\pm$0.14 & 5.72 & 5.52$\pm$0.09 & 3151$\pm$36 \\ 
102756832 &  UNC & 101.2318 & -2.0673 & M5 & III & LPV? & 16.57 & 14.61$\pm$0.22 & 12.10 & 10.27$\pm$0.17 & 8.71 & 8.19$\pm$0.11 & 6.95 & 6.73$\pm$0.07 & 3189$\pm$45 \\ 
102798403 &  UNC & 101.4759 & -2.3579 & M2 & I & LPV? & 13.96 & 11.68$\pm$0.35 & 11.97 & 10.29$\pm$0.28 & 10.61 & 10.00$\pm$0.18 & 9.74 & 9.48$\pm$0.12 & 4873$\pm$318 \\ 
102862904 &  UNC & 101.9196 & -2.4779 & M2 & III & LPV? & 16.85 & 14.05$\pm$0.24 & 13.69 & 11.57$\pm$0.18 & 10.89 & 10.15$\pm$0.11 & 9.18 & 8.86$\pm$0.08 & 3538$\pm$63 \\ 
102868800 &  UNC & 101.9532 & -2.0940 & M2 & III & LPV? & 16.87 & 13.96$\pm$0.22 & 13.36 & 11.15$\pm$0.17 & 10.31 & 9.54$\pm$0.11 & 8.47 & 8.14$\pm$0.07 & 3417$\pm$47 \\ 
102885433 &  DJM  47 & 102.0524 & -0.9607 & M1 & III & C* & 14.56 & 12.94$\pm$0.19 & 10.76 & 9.35$\pm$0.15 & 7.80 & 7.37$\pm$0.10 & 5.46 & 5.28$\pm$0.06 & 3205$\pm$45 \\ 
102921590 &  UNC & 102.2414 & -1.5149 & M2 & III & LPV? & 13.58 & 12.00$\pm$0.22 & 11.47 & 10.26$\pm$0.18 & 9.50 & 9.08$\pm$0.11 & 8.18 & 8.00$\pm$0.08 & 3878$\pm$92 \\ 
102930515 &  UNC & 102.2867 & -1.1656 & M0 & III & LPV? & 13.48 & 11.75$\pm$0.19 & 11.94 & 10.71$\pm$0.15 & 10.54 & 10.08$\pm$0.10 & 9.59 & 9.40$\pm$0.07 & 4752$\pm$187 \\ 
102994604 &   IRAS 06480-0305 & 102.6317 & -3.1452 & M6 & III &  LPV* & 16.04 & 14.08$\pm$0.19 & 11.44 & 9.58$\pm$0.14 & 7.97 & 7.45$\pm$0.09 & 6.26 & 6.04$\pm$0.06 & 3179$\pm$35 \\ 
\hline
\hline
\end{tabular}
}
\end{table*}

\end{document}